# On the Scalability of Big Data Cyber Security Analytics Systems

Faheem Ullah[†], M. Ali Babar
CREST – Centre for Research on Engineering Software Technologies, The University of Adelaide, Australia
Email: {faheem.ullah, ali.babar}@adelaide.edu.au

**Abstract**

<u>B</u>ig <u>D</u>ata <u>C</u>yber Security <u>A</u>nalytics (BDCA) systems use big data technologies (e.g., Apache Spark) to collect, store, and analyse a large volume of security event data for detecting cyber-attacks. The volume of digital data in general and security event data in specific is increasing exponentially. The velocity with which security event data is generated and fed into a BDCA system is unpredictable. Therefore, a BDCA system should be highly scalable to deal with the unpredictable increase/decrease in the velocity of security event data. However, there has been little effort to investigate the scalability of BDCA systems to identify and exploit the sources of scalability improvement. In this paper, we first investigate the scalability of a Spark-based BDCA system with default Spark settings. We then identify Spark configuration parameters (e.g., execution memory) that can significantly impact the scalability of a BDCA system. Based on the identified parameters, we finally propose a parameter-driven adaptation approach, *SCALER*, for optimizing a system's scalability. We have conducted a set of experiments by implementing a Spark-based BDCA system on a large-scale OpenStack cluster. We ran our experiments with four security datasets. We have found that (i) a BDCA system with default settings of Spark configuration parameters deviates from ideal scalability by 59.5% (ii) 9 out of 11 studied Spark configuration parameters significantly impact scalability and (iii) *SCALER* improves the BDCA system's scalability by 20.8% compared to the scalability with default Spark parameter setting. The findings of our study highlight the importance of exploring the parameter space of the underlying big data framework (e.g., Apache Spark) for scalable cyber security analytics.

## 1. Introduction

The volume and velocity of digital data are increasing enormously. The amount of digital data increased to 40 trillion gigabytes in 2020, which was merely 1.2 trillion gigabytes back in 2010 [1]. Given that "*data is the new oil of the digital economy*", the amount of data analysed has jumped from 0.5% in 2012 to 37% in 2019 [1, 2]. However, the traditional software systems (e.g., relational database and data warehouse) are unable to collect, store, and analyse such a large volume of data. Therefore, big data storage and processing technologies (e.g., Hadoop, Spark, and Cassandra) are being increasingly leveraged in various fields to deal with the massive volume, velocity, and variety of data such is evident from the fact that 97.2% organizations are investing in big data [3, 4]. For instance, Persico et al. [5] compared two big data architectures (Lambda and Kappa) using Microsoft Azure cloud platform for social network data analysis. In another study [6], Aceto et al. leveraged big data technologies for implementing deep learning to classify encrypted mobile traffic. In the healthcare domain, several studies (e.g., [7-9]) explored the incorporation of big data technologies for medical image analysis, genomic analysis, and prediction of various diseases. Similarly, big data technologies are increasingly used in the oil and gas sector for analysing the large volume, velocity, and variety of data related to drilling, exploration, and production [10, 11].

Similar to the other domains such as bioinformatics [12] and healthcare [13], the role of big data technologies is on the rise in the cyber security domain too. The significance of big data technologies in the cyber security domain was first highlighted by Cloud Security Alliance (CSA) in 2013 [14]. A CSA report emphasizes the important need of enabling the traditional cyber security systems (e.g., intrusion detection system and malware detection system) to deal with the massive volume, velocity, and variety of security event data such as NetFlow, Firewall logs, and Packet data [14]. The merger of cyber security systems and big data technologies has given birth to a new breed of a software system called **B**ig **D**ata **C**yber Security **A**nalytics (BDCA) system, which is defined as

---

[†] Corresponding Author

"*A system that leverages big data technologies for collecting, storing, and analyzing a large volume of security event data to protect organizational networks, computers, and data from unauthorized access, damage, or attack*" [15]. A recent study of BDCA systems indicates that 72% of organizations that employed big data technologies in their cyber security landscape reported significant improvement in their cyber agility [16].

BDCA systems are primarily classified into two categories based on their attack detection capability – Generic BDCA systems and Specific BDCA systems [15]. *Generic BDCA systems* (e.g., an intrusion detection system supported with big data technologies) aim to detect a variety of attacks such as SQL injection, cross-site scripting, and brute force. *Specific BDCA systems* (e.g., a phishing detection system built using big data technologies) are focused on detecting a specific attack type such as phishing. The main characteristics of BDCA systems, that distinguishes them from traditional cyber security systems, include (i) monitoring diverse assets of an enterprise such as data storage systems, computing machines, and end-user applications (ii) integrating security data from multiple sources such as IDS, firewall, and anti-virus (iii) analysing large volume of security event data in near real-time (iv) enabling deep and holistic security analytics for unfolding complex attacks such as Advanced Persistent Threats (APT) and (v) analysing heterogeneous streams of security event data [15].

Like any software system, certain quality attributes (e.g., interoperability and reliability) are expected in a BDCA system. Ullah and Ali Babar [15] reported the 12 most important quality attributes of a BDCA system, where scalability is ranked as the third most important quality attribute of a BDCA system. Scalability is defined as "*the system's ability to increase speed-up as the number of processors increase*" [17]. The rationale behind the need for a BDCA system being highly scalable is twofold – (a) the volume of security event data is rapidly increasing, which requires a BDCA system to scale up (by adding more computational power) to process data without impacting the response time of a system [18, 19] and (b) the velocity of security event data generation fluctuates [20, 21]. For example, a BDCA system analysing network traffic of a bank experiences a higher workload during working hours as compared to non-working hours. Therefore, a BDCA system should efficiently use the commodity or third-party resources to scale up during working hours and scale down during non-working hours. In other words, a BDCA system should take maximum benefit from the additional resources.

Among the 74 studies on BDCA reviewed in [15], 40 studies highlight the importance of scalability for a BDCA system. However, none of the studies have either investigated the factors that impact the scalability of a BDCA system or have proposed any solutions for improving scalability. Several BDCA studies (e.g., [22-24]) hint at factors such as machine learning algorithm employed in a system, quality of security event data, and big data processing framework that can potentially impact scalability. Among these factors, the most prominent is the underlying big data processing framework such as Spark or Hadoop, which is an integral part of any BDCA system. One of the core features of any big data processing framework is its configuration parameters (e.g., executor memory) [25], which guide how a framework should process data. For example, executor memory specifies how much memory should be allocated to an executor process. The importance of parameter configuration for big data processing frameworks has been highlighted by several studies (e.g., [18, 26, 27]). However, none of the previous studies have investigated their impact on the scalability of a big data system. Therefore, this paper aims to "***investigate the impact of Spark configuration parameters on the scalability of a BDCA system and devise an approach for improving the scalability***". Given that there exist several big data processing frameworks (e.g., Spark [25], Hadoop [28], Storm [29], Samza [30], and Flink [31]), we investigate Spark as it is currently the most widely used framework in the domain of BDCA. We have observed that 14 BDCA studies published in 2014 used Hadoop and only four used Spark which changed to four studies using Hadoop and five studies using Spark in 2017 [15]. A similar dominance of Spark over Hadoop in the industry is observed [32]. To achieve the aforementioned aim, this paper contributes to the state-of-the-art by answering the following Research Questions (RQ).

> *RQ1: How does a BDCA system scales with default Spark configuration settings?*
> *RQ2: What is the impact of tuning Spark configuration parameters on the scalability of a BDCA system?*
> *RQ3: How to improve the scalability of a BDCA system?*

To answer the three research questions, we developed an experimental infrastructure on a large-scale OpenStack cloud. We implemented a Spark-based BDCA system that ran on an OpenStack cloud in a fully distributed fashion. We used two evaluation metrics – the accuracy and scalability of a BDCA system. For measuring accuracy, we leveraged the commonly used measures such as F1 score, precision, and recall. For measuring scalability, we used the scalability scoring measure reported in Section 0. We used four security datasets (i.e., KDD [33], DARPA [34], CIDDS [35], and CICIDS2017 [36]) in our experimentation and evaluated the BDCA system with four learning algorithms (i.e., Naïve Bayes, Random Forest, Support Vector Machine, and Multilayer Perceptron) that are employed in the system for classifying security data into benign and malicious categories. Based on our comprehensive experimentation, we have found that:

(i) A BDCA system with default Spark configuration parameters does not scale ideally. The deviation from ideal scalability is around 59.5%. This means a system only takes 41.5% benefit from the additional resources.
(ii) Among the 11 investigated Spark parameters, changing the value of nine parameters significantly impacts a BDCA system's scalability. The optimal value of a parameter (with respect to scalability) varies from dataset to dataset.
(iii) We proposed and evaluated a parameter-driven adaptation approach, *SCALER*, that automatically selects the most optimal value for each parameter at runtime. The evaluation results show that on average, *SCALER* improves a BDCA system's scalability by 20.8%.

The rest of this paper is structured as follows. Section 2 reports the security datasets, our BDCA system, the instrumentation setup, and evaluation metrics. Our adaptation approach is presented in Section 3. Section 4 presents the detailed findings of our study with respect to the three research questions. Section 5 presents our reflections on the findings. Section 6 positions the novelty of our work with respect to the related work. Finally, Section 7 concludes the paper by highlighting the implications of our study for practitioners and researchers.

## 2. Research Methodology

This section describes the datasets, our BDCA system, the instrumentation setup, and evaluation metrics.

### 2.1. Security Datasets

In order to answer the three research questions (Section 1), we used four security datasets: KDD [33], DARPA [34], CIDDS [35], and CICIDS2017 [36]. These datasets are briefly described in the following with their details presented in Table 1. We selected these four datasets as they vary from each other in terms of attack types, the number of training and testing instances, dataset size, publication dates, redundancy, and the number of features (e.g., source IP, source port, and payload). These characteristics of the selected datasets are expected to provide rigour and generalization to our findings. It is important to note that we used the whole of these datasets, instead of using a small sample of each dataset, in our experiments.

***KDD:*** The KDD dataset contains 494,022 records as training data and 292,300 records as testing data. Each record represents a network connection – consisting of 41 features. Each record is labelled as belonging to either the normal class or one of the four attack classes, i.e., Denial of Service, Probing, Remote to Local, and User to Root. The testing data includes attack types that are not present in the training data, which makes the evaluation more realistic. More details on the dataset are available in [33].

***DARPA:*** Similar to KDD, the records in this dataset are divided into training and testing subsets. The training data consists of 2,723,496 records, while the testing data consists of 1,522,310 records. Each record represents a network connection – consisting of six features. Each record is labelled as 0 or 1, where 0 specifies a normal connection and 1 specifies an attack. The attack types present in DARPA are the same as KDD. More details on the DARPA dataset are available in [34].

***CIDDS:*** This dataset has been recently developed as KDD and DARPA are relatively old datasets. The CIDDS dataset consists of four-week NetFlow data directed towards two servers, i.e., OpenStack and External Server.

Table 1: Number of training and testing instances in each dataset

| Dataset | Number of Features | No. of Instances in Training Dataset | No. of Instances in Testing Dataset |
|---|---|---|---|
| KDD | 41 | 494,022 | 292,300 |
| DARPA | 6 | 2,723,496 | 1,522,310 |
| CIDDS | 9 | 5,634,347 | 2,788,463 |
| CICIDS2017 | 77 | 1,311,822 | 445,061 |

The training dataset contains 5,634,347 records and the testing dataset contains 2,788,463 records. Each record represents a network connection – consisting of nine features. The dataset contains four types of attacks: *pingScan*, *portScan*, *bruteForce*, and *DoS*. More details on the dataset are available in [35].

***CICIDS2017:*** This is also a recently developed dataset, which contains a variety of state-of-the-art attacks. The dataset consists of five days of network traffic directed towards a network consisting of three servers, a firewall, a switch, and 10 PCs. The training dataset consists of 1,311,822 records and the testing dataset consists of 445,061 records. Each record consists of 77 features. This dataset contains six types of attacks: *bruteForce*, *heartBleed*, *botNet*, *DoS*, *Distributed DoS*, *webAttack*, and *infiltration attack* [36].

### 2.2. Our BDCA System

The overview of our BDCA system is depicted in Figure 1. This system consists of three layers –*Security Analytics Layer*, *Big Data Support Layer*, and *Adaptation Layer*. In the following, we describe Security Analytics Layer and Big Data Support Layer, while the details of the Adaptation Layer are presented in Section 3.

#### 2.2.1. Security Analytics Layer

This layer processes the security event data for detecting cyber-attacks. The layer consists of three phases (i.e., data engineering, feature engineering, and data processing), which are described below.

***Data Engineering:*** This phase pre-processes the data to handle missing values and remove incorrect values and outliers [15]. A negative value indicates that the number of features for the instances is incomplete, hence, the instance is removed. Incorrect values (e.g., standard deviation = -1) in the dataset that are unacceptable for the Machine Learning (ML) model employed in a system for the classification of security event data into normal and attack classes. Therefore, we use the filter method of *DataFrame* available in the Spark package, i.e., *org.apache.spark.sql* [37] to remove the incorrect values. The existence of the outliers in the training dataset affects the accuracy of the machine learning model [38]. We, therefore, removed the values that were larger than *Double.MaxValue*. CICIDS2017 has missing values, therefore, we removed the instances with the missing values by simply investigating whether the value of the last feature is negative.

***Feature Engineering:*** This phase generates new features and/or transforms the values of features into a new range [15]. For all four datasets, we assembled the features to transform multiple columns of features into one column of feature vector for fitting the ML model. We used *VectorAssembler* method in *org.apache.spark.ml.feature* for the implementation of assembling the features. Since some algorithms (e.g., Naïve Bayes) in SparkML library cannot handle non-numeric features, we used *StringIndexer* (from *org.apache.spark.ml.feature*) to transform the label features (i.e., normal and attack) in the KDD dataset from string to indices. Given the relatively smaller number of features in the DARPA dataset, we expanded the features to a polynomial space. We used *PolynomialExpansion* method in *org.apache.spark.ml.feature* for feature expansion.

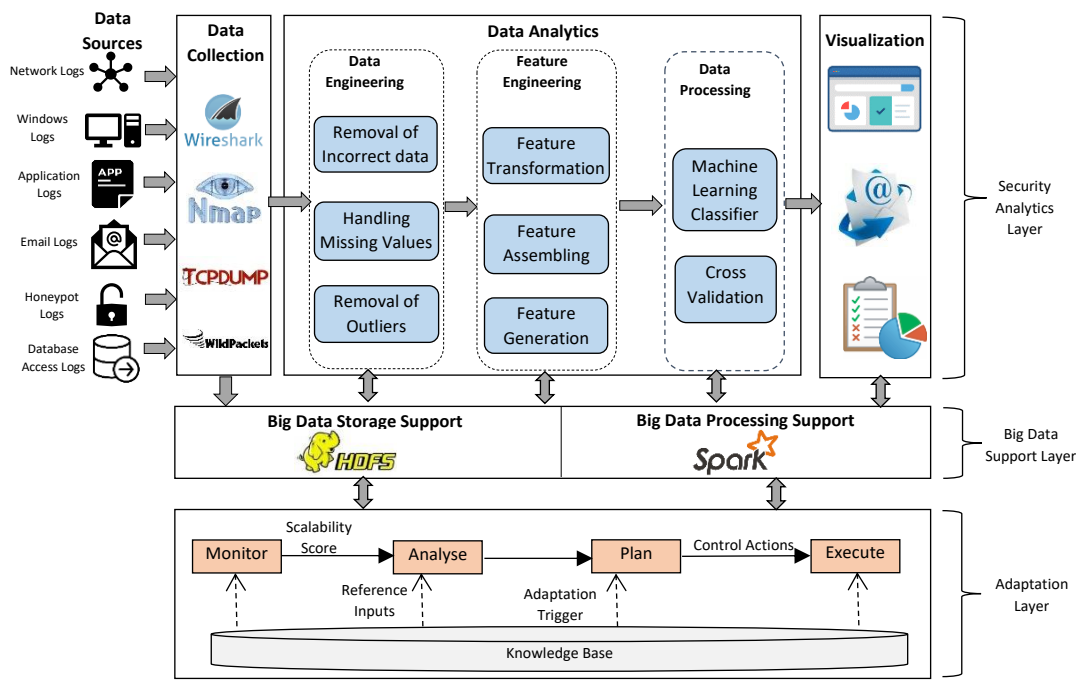

Figure 1: An overview of our BDCA system

***Data Processing:*** This phase leverages ML/DL algorithm to classify the instances in security data as either *normal* or *attack*. In our system, we separately used four ML/DL algorithms - Naïve Bayes (NB), Random Forest (RF), and Support Vector Machine (SVM), for classifying the instances. These four algorithms have been selected based on (i) their widespread use in the domain of BDCA [15] (ii) popularity in Kaggle competitions and (iii) availability in Spark ML library and DeepLearning4j [39]. We used Spark package *org.apache.spark.ml.classification* for implementing the ML algorithms. For cross-validation of ML models, we used *CrossValidator* method available in *org.apache.spark.ml.tuning*.

2.2.2. Big Data Support Layer

This layer manages the distributed storage and processing of data on multiple computing nodes. The layer consists of big data processing framework (i.e., Spark) and big data storage (i.e., HDFS). Apache Spark is an open-source big data processing framework that uses in-memory primitives to process a large amount of data. Spark is quite suitable for ML tasks, which requires iterative processing that best suits Spark architecture [37]. Moreover, Spark is not only much faster than Hadoop, but is also compatible with multiple file systems such as HDFS, MongoDB, and Cassandra. Hadoop Distributed File System (HDFS) is a data storage system that enables the distributed storage of a massive amount of data [40]. By default, HDFS replicates each block of data on three nodes, which makes it quite fault-tolerant.

2.3. Instrumentation Setup

We configured Spark and Hadoop (for HDFS) on an OpenStack cluster consisting of 10 computing nodes. Each node is installed with Ubuntu 16.04 Xenial Xerus operating system. Each node runs Spark 2.4.0, Hadoop 2.9.2, and JDK 1.8. The 10 computing nodes are divided into master and slave nodes. There is one master with m1.large flavour (8 GB RAM, 80 GB Hard disk, and 8 virtual CPUs) and nine worker nodes with m1.small flavour (2 GB RAM, 10 GB Hard disk, and one virtual CPU). Each node in the cluster has a floating IP for communicating with the external world and an internal IP for communicating with other nodes in the cluster. To associate floating IP with internal IP, a router is created as the bridge between the external network (floating IPs) and subnets (internal IPs). We used Scala programming language for various implementations on Spark.

2.4. Evaluation Metrics

In this study, we assess two qualities of our BDCA system – *accuracy* and *scalability*. Accuracy measures how accurately a BDCA system classifies the instances in the datasets into normal and attack categories. Scalability measures to what extent our system takes advantage of the additional hardware resource added to a system in the form of computing nodes.

2.4.1. Measuring Accuracy

For assessing accuracy, we used five evaluation metrics that are commonly used in the BDCA domain [15]. These metrics include False Positive Rate, F-Score, Recall, Accuracy, and Precision. Table 2 provides the definition and brief description of each of the metrics.

Table 2: Evaluation metrics for assessing accuracy and their descriptions. TP – Ture Positive, FP – False Positive, TN – True Negative, and FN – False Negative

| Metric | Definition | Description |
| --- | --- | --- |
| Precision | $P = \frac{TP}{TP + FP}$ | Proportion of instances correctly classified as attack instances |
| Recall | $R = \frac{TP}{TP + FN}$ | Proportion of attack instances that are correctly classified |
| F-score | $F = 2 \times \frac{P \times R}{P + R}$ | Harmonic mean of precision and recall |
| Accuracy | $A = \frac{TP + TN}{TP + TN + FP + FN}$ | Proportion of correctly classified instances |
| False Positive Rate | $FPR = \frac{FP}{FP + TN}$ | Proportion of normal instances classified as attack instances |

### 2.4.2. Measuring Scalability

Several studies (e.g., [41-43]) have proposed metrics for measuring the scalability of a system. However, the previous metrics are not suitable for the scalability analysis in our study for two reasons: 1) these metrics do not quantify scalability with respect to ideal scalability, which is required for evaluating the effectiveness of our adaptation approach presented in Section 3; 2) these metrics are primarily suitable for measuring scalability in cases where a system is partly executed in parallel mode and partly in sequential mode whereas our system (implemented using Apache Spark [25, 44]) is executed fully in parallel mode. For this study, we used Eq. 1 to measure the scalability of a BDCA system. In Eq. 1, $S(c)$ denotes the scalability score for curve '$c$'. *Gap* denotes the quantified gap value between the achieved and ideal response time (i.e., training time or testing time), which is calculated using Eq. 2. In Eq. 2, $\omega_n$ represents the user-defined weight that specifies the importance of the gap between achieved and ideal response time at '$n$' worker nodes. For example, $\omega_2$ is the weight for specifying the importance of gap at two worker nodes and $\omega_4$ is the weight for specifying the importance of the gap at four worker nodes. In Eq. 1, $\omega_{n+1}$ denotes the weight for specifying the importance of gap at one size larger than the existing cluster size, e.g., to specify the importance of gap beyond eight nodes if n (cluster size) equals to eight. The sum of all weights is equal to 1 (as presented in Eq. 5). In Eq. 2, $G_n$ defines the ratio of unaccomplished response time improvement to the response time improvement in the ideal case with '$n$' worker nodes. $G_n$ is calculated using Eq. 3, where $AT_n$ denotes achieved response time with '$n$' worker nodes and $IT_n$ denotes the ideal response time with '$n$' worker nodes. In Eq. 1, *Trend*, which is calculated using Eq. 4, denotes how response time decreases between the last two cluster setups such as from six to eight worker nodes in with cluster of size 8, the higher of which indicates the probability that the response time tends to decrease with more than eight nodes.

$$S(c) = 1 - Gap - \omega_{n+1} \times (1 - Trend) \quad (1)$$

$$Gap = \sum_{i=1}^{n} \omega_{2i} G_{2i} \quad (2)$$

$$G_n = \frac{AT_n - IT_n}{IT_1 - IT_n} \quad (3)$$

$$Trend = \frac{AT_{n-1} - AT_n}{IT_{n-1} - IT_n} \quad (4)$$

$$\sum_{i=1}^{n} \omega_{2i} = 1 \quad (5)$$

***Example Scenario:*** We illustrate the use of the scalability metric with an example, which includes eight hypothetical scalability scenarios for a software system. Table 3 presents the hypothetical response times for eight different scenarios with respect to five different cluster configurations, i.e., 1 worker, 2 workers, 4 workers, 6 workers, and 8 workers. Figure 2 shows the eight scalability curves drawn using the response times reported in Table 3. The eight scenarios (i.e., ideal scenario and Scenario 1 – Scenario 7) presented in Table 3 and Figure 2 differ from each other with respect to two parameters – the number of worker nodes and response time. The number of worker nodes is the independent parameter that we change to observe the impact on the dependent parameter i.e., response time. As shown in Figure 2, the impact of change in the number of worker nodes is not consistent across scenarios. This could possibly be due to multiple reasons in a real-world scenario. For example, in *Ideal Scenario*, the system utilizes the underlying resources such as CPU and RAM more efficiently as compared to *Scenario-1*. Therefore, the response time in the *Ideal Scenario* reduces more significantly with the increase in the number of worker nodes as compared to the reduction in response time in *Scenario-2*.

*Ideal Scenario* underlines the case where each time the number of nodes is doubled, the response time is reduced to half. For calculating the scalability score, we use a value of 0.2 for all weights (e.g., $\omega_2, \omega_4, \omega_6, \omega_8, \omega_{10}$). For calculating Trend in this scenario, $AT_6 = 1.33$, $AT_8 = 1$, $IT_6 = 1.33$, and $IT_8 = 1$ as shown in Table 3, hence, Trend = 1 using Eq. 4. Since there is no gap between achieved and ideal response time, the value of all gaps is equal to zero (i.e., $G_2 = 0$, $G_4 = 0$, $G_6 = 0$, and $G_8 = 0$). Thus, the overall gap is zero (i.e., Gap = 0) calculated using Eq. 3. Feeding these values into Eq. 1 gives us S(ideal) = **1.00**. For *Scenario-1*, $AT_6 = 4$, $AT_8 = 3$, $IT_6 = 1.66$, and $IT_8 = 1.25$, hence, Trend = 2.44, which is high – indicating a positive trend of scalability after eight worker nodes. The values of $G_n$ is calculated using Eq. 3 are $G_2 = 1.2$, $G_4 = 0.46$, $G_3 = 0.28$, and $G_8 = 0.2$, which gives Gap = 0.42. Hence, the scalability score for *Scenario-1* is **0.85**, which indicates poor scalability as compared to ideal scalability. This is also observable from the comparison of the two curves, i.e., *Ideal Scenario* and *Scenario-1* as depicted in Figure 2. As compared to *Scenario-1*, there is a higher reduction trend in response time with the increase in the number of nodes in the ideal case. Thus, the scalability score of *Scenario-1* is smaller as compared to the ideal scenario.

Table 3: Response time (in sec) and scalability score for the eight hypothetical scalability scenarios

| Number of Worker Nodes | Response Time (sec) | | | | | | | |
|---|---|---|---|---|---|---|---|---|
| | Ideal Scenario | Scenario-1 | Scenario-2 | Scenario-3 | Scenario-4 | Scenario-5 | Scenario-6 | Scenario-7 |
| 1 | 8.00 | 10.00 | 10.00 | 8.00 | 8.00 | 9.50 | 8.00 | 8.00 |
| 2 | 4.00 | 11.00 | 6.87 | 5.50 | 7.00 | 5.00 | 7.00 | 8.00 |
| 4 | 2.00 | 6.00 | 5.00 | 4.00 | 6.00 | 8.00 | 7.20 | 8.00 |
| 6 | 1.33 | 4.00 | 3.75 | 3.00 | 5.80 | 5.50 | 6.00 | 8.00 |
| 8 | 1.00 | 3.00 | 3.20 | 2.90 | 5.70 | 6.00 | 7.20 | 8.00 |
| Scalability Score | 1.00 | 0.85 | 0.83 | 0.61 | 0.31 | 0.16 | -0.56 | 0.00 |

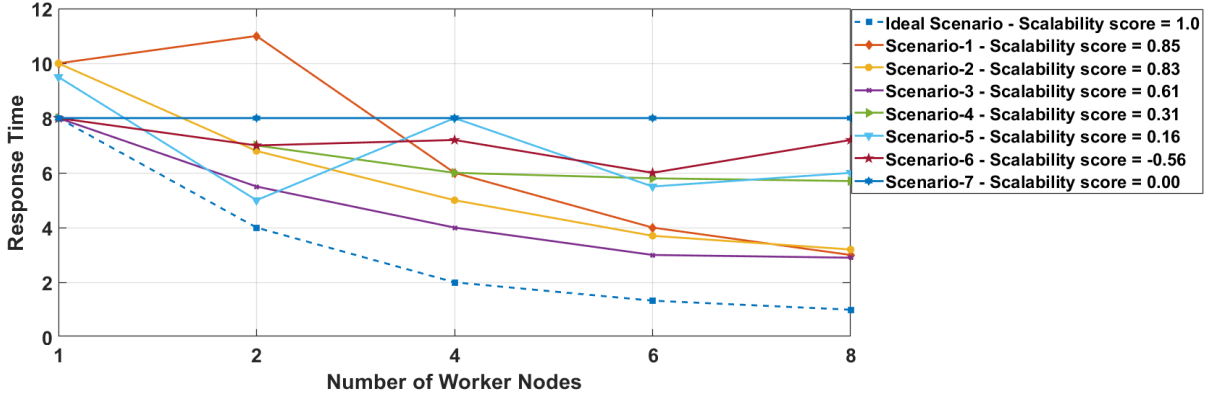

Figure 2: Hypothetical scalability scenarios (drawn based on Table 3) to illustrate the use of the scalability metric

The scalability score for *Scenario-2* is **0.83**, which is slightly lower than the scalability score for *Scenario-1*. The slight difference is mainly due to the difference in Trend for the two scenarios, i.e., a reduction from 4 to 3 in *Scenario-1* and a reduction from 3.75 to 3.2 in *Scenario-2*. The scalability score for *Scenario-4* is **0.31**, which is quite lower as compared to *Scenario-1* and *Scenario-2*. If we observe the scalability curve for *Scenario-4* in Figure 2, the response time reduces quite significantly as we increase the number of worker nodes from 1 to 4 nodes. However, there is almost no reduction as the number of worker nodes are increased from 4 to 8, which is why the scalability score is much lower as compared to more smoother curves such as the curves for *Scenario-1* and *Scenario-2*. In *Scenario-5*, the sudden upward jump in the curve from 2 nodes to 4 nodes impacts the scalability score of the whole curve. Therefore, the scalability is quite low, i.e., **0.16**. In *Scenario-6*, the response time increases (unlike as expected) at two transitions, i.e., from 2 nodes to 4 nodes and from 6 nodes to 8 nodes. The spike in response time from 6 to 8 nodes is quite high. Therefore, the negative impact on the response time at two transitions significantly impacts the scalability score and Eq. 1 generates a much lower scalability score (i.e., **-0.56**) for *Scenario-6*. The response time in *Scenario-7* does not change with the change in the number of nodes, therefore, the scalability score for *Scenario-7* is **0.00**.

## 3. Our Adaptation Approach

To optimize the scalability of a BDCA system, we present *SCALER* - an adaptation approach that automatically triggers the tuning process and tune Spark configuration parameters. By tuning, we mean to select a combination of parameters, which generates a scalability score that is above the predefined threshold (Section 3.3).

Spark parameters control most of the application settings and directly impact the way an application runs [45]. All Spark parameters have a default configuration; however, the default configuration is not suitable for each application [45]. Therefore, the parameters need to be configured separately for each application. The spark parameters investigated in this study for their impact on the scalability of a system are presented in Table 4. We selected 11 parameters based on the following criteria – (i) the parameters have proven impact on different aspects of Spark such as scheduling, compression, and serialization (ii) the parameters contribute to Spark running time as highlighted in [46] and [47] and (iii) the parameters impact multiple levels (e.g., machine level and cluster level) of a BDCA system as reported through industry practices [32, 45]. Although *SCALER* considers 11 Spark parameters, it is worth noting that *SCALER* can be easily extended to incorporate more parameters if needed. In the following, we describe our adaptation approach that automatically tunes Spark configuration parameters for improving scalability. We present our adaptation approach as per the guidelines for adaptation approaches presented by Villegas et al. [48].

Table 4: Spark parameters considered in this study for their scalability impact and subsequent tuning

| ID | Spark Parameters | Default Value | Description |
|---|---|---|---|
| P1 | Spark.executor.memory | 1024 MB | Amount of memory used by each executor process in Spark |
| P2 | Spark.shuffle.sort.bypassMergeThreshold | 200 | Underlines the threshold of reduce partitions for avoiding merge-sorting data |
| P3 | Spark.shuffle.compress | TRUE | Whether (or not) to compress the map output files |
| P4 | Spark.memory.storageFraction | 0.5 | Amount of memory available for task execution |
| P5 | Spark.shuffle.file.buffer | 32 KB | Amount of memory available to buffer file output streams |
| P6 | Spark.reducer.maxSizeInFlight | 48 m | Maximum size of map output to be fetched for reducer |
| P7 | Spark.memory.fraction | 0.6 | Proportion of heap size used for execution and storage |
| P8 | Spark.serializer.objectStreamReset | 100 | To allow or stop garbage collection of objects |
| P9 | Spark.rdd.compress | FALSE | Whether (or not) to compress Resilient Data Distributed Datasets (RDD) |
| P10 | Spark.shuffle.memoryFraction | \(Deprecated) | Proportion of Java heap size used for aggregation. Beyond this limit, contents start spelling to the disk |
| P11 | Spark.driver.memory | 1024 MB | Amount of memory used for initializing Spark context |

### 3.1. Adaptation Goal

The adaptation goal is stimulated by the main reason for adaptation, i.e., why a BDCA system needs to adapt [48]. Our adaptation goal is driven by the results collected for RQ1 (Section 4.1) and RQ2 (Section 4.2). These findings indicate that (i) with default settings of Spark parameters, a BDCA system does not scale ideally and (ii) 9 out of 11 studied spark parameters significantly impact the scalability of a BDCA system. Therefore, our adaptation approach aims to *"automatically tune Spark parameters for improving the scalability of a BDCA system"*.

### 3.2. Reference Inputs and Measured Outputs

Reference inputs delineate the target to be achieved through adaptation [48]. For our approach, ideal scalability is the reference input. Ideal scalability implies that when the number of computing nodes is doubled, the job completion time (i.e., data processing time) is reduced by half. As illustrated in Section 2.4.2, the scalability score for ideal scalability is **1.0**, which *SCALER* aims to achieve. Whilst reference inputs are specified by a user, the measured outputs are the actual measures collected from a running system [48]. The measured outputs are then compared with the reference inputs to assess the extent to which a system has achieved the target state (indicated by reference input) through adaptation. For our adaptation approach, the measured outputs are the scalability scores calculated when a system is under operation. The scalability score is then compared with the scalability score for the ideal scalability (i.e., 1.0) to assess the extent to which our adaptation approach has achieved its target.

### 3.3. Adaptation Trigger

An adaptation trigger defines the condition for triggering the adaptation process [48]. In our approach, we define a threshold value for the scalability score. Our approach constantly monitors the scalability score of a system and whenever the scalability score is less than the threshold value, the adaptation is triggered to optimize the scalability score. In Algorithm 1, line 5 specifies the adaptation trigger condition. In order to determine the threshold value for the evaluation of our approach, we first calculated the scalability score for the 108 use cases (i.e., 9 parameter configurations × 4 datasets × 4 algorithms) reported in Section 4.2 using Eq. 1 to Eq. 5. For calculating the scalability scores, we used the same value (i.e., 0.2) for all weights (e.g., $\omega_2, \omega_4, \omega_6, \omega_8, \omega_{10}$) in Eq. 1 to Eq. 5 to specify the same level of importance for all the gaps between ideal and achieved response time. We computed the mean scalability score of the 108 use cases, which gives a value of 0.55. Although the value 0.55 underlines the mean scalability score for the considered scenario that can be set as a threshold, we increased the threshold

**Algorithm. 1.** Algorithm for adapting configuration setting of Spark-based BDCA system

| | | |
|---|---|---|
| **Inputs:** | $S_{Threshold}$ = Threshold Scalability Score | |
| | $S_{Increment}$ = Increment Scalability Threshold | |
| | CPV = {$CPV_1$, $CPV_2$, …, $CPV_n$} | // *Set of Combination of Parameter Values* |
| | $S_{optimal}$ = The most optimal scalability score | |
| | $R_t$ = The number of times each execution is repeated | |
| **Output:** | $CPV_{optimal}$ = CPV with scalability score above $S_{Threshold}$ | |
| **Steps:** | 1. $CPV_{optimal} \leftarrow CPV_{default}$    // $CPV_{default}$ = {A, A, A, A, A, A, A, A, A} | |
| | 2. $S_{default} \leftarrow 0$, $S_{optimal} \leftarrow 0$, $R_t \leftarrow 3$, $S_{Threshold} \leftarrow 0.58$, $S_{Increment} \leftarrow 0.036$    // *Initialize the variables* | |
| | 3. runBDCA with $CPV_{default}$    // *execute the BDCA system with default Spark configuration settings* | |
| | 4. Calculate Scalability Score $S_{default}$ of the BDCA system using Eq. 1 | |
| | 5.    **if** $S_{default} > S_{Threshold}$ **then** | |
| | 6.        $CPV_{optimal} \leftarrow CPV_{default}$ | |
| | 7.        return $CPV_{optimal}$ | |
| | 8.    **end if** | |
| | 9.    **for** i = 1 to CPV.legnth **do** | |
| | 10.        CPV [i] ← B    //*change the value of parameter from A to B as mentioned in Table 5* | |
| | 11.        **for** j = 1 to $R_t$ **do** | |
| | 12.            runBDCA with 1 worker node | |
| | 13.            runBDCA with n-1 worker nodes    // *n specifies cluster size (total number of nodes)* | |
| | 14.            runBDCA with n worker nodes | |
| | 15.        **end for** | |
| | 16.        Calculate Trend using Eq. 4    // *To investigate scalability trend from 6 to 8 nodes* | |
| | 17.        **if** Trend < 0 **then** | |
| | 18.            CPV [i] ← A | |
| | 19.            **continue** | |
| | 20.        **end if** | |
| | 21.        **for** k = 1 to $R_t$ **do** | |
| | 22.            runBDCA with 'n-6' worker nodes | |
| | 23.            runBDCA with 'n-4' worker nodes | |
| | 24.        **end for** | |
| | 25.        Calculate Scalability Score S[i] for CPV [i] using Eq. 1 | |
| | 26.        **if** S[i] > $S_{optimal}$ **then** | |
| | 27.            $S_{optimal}$ ← S[i] | |
| | 28.            $CPV_{optimal}$ ← CPV[i] | |
| | 29.            **if** S[i] > $S_{Threshold}$ **then** | |
| | 30.                **break** | |
| | 31.            **end if** | |
| | 32.        **end if** | |
| | 33.    **end for** | |
| | 34. return $CPV_{optimal}$ | |

value by setting the value of *incrementthreshold* equal to 0.0365. This is because setting *incrementthreshold* to a positive value makes the evaluation of *SCALER* more robust as reported in Section 4. We selected the value 0.0365 based on the fact that our threshold value should be higher than both mean and median. Hence, adding 0.0365 to the mean scalability score gives us a value of 0.58, which is equal to the median value too. As a result, the final threshold value for triggering adaptation is **0.58**. Whilst we selected the value of *incrementthreshold* keeping in view the median value of the 108 cases, the main objective of having the *incrementthreshold* is to render flexibility to users in terms of how robust the user wants the system to be in terms of adaptation.

### 3.4. Control Actions

Once adaptation is triggered, the control actions are automatically taken by our adaptation approach to adapt a system. The adaptation process is only triggered when a system's scalability score gets below the threshold scalability score and so the control actions aim to make the scalability score above the threshold. In our approach, the controls actions execute a system with different Spark parameter configurations with the aim to find a combination of parameters with which a system's scalability score gets above the threshold scalability score. The control actions only try changing the parameters that have a significant impact on scalability as determined in RQ2 (Section 4.2). These parameters with a significant impact on scalability are presented in Table 5 with the default and modified value for each parameter. With respect to value options, there are two types of parameters - Boolean parameters and numerical parameters. A Boolean parameter takes only two values (i.e., TRUE and FALSE). The values options for numerical parameters presented in Table 5 are chosen based on academic and industrial recommendations [32, 45]. For example, the execution memory can be set as 1024m (default value) or 1250m (modified value). In Table 5, 'A' represents the default value and 'B' represents the modified value for a parameter. For instance, P1-A and P1-B are default and modified values for parameter P1 (*Spark.executor.memory*). Some sample Combinations of Parameter Values (CPV) are shown in Table 6.

Table 5: Impactful Spark parameters used in the adaptation approach and their potential value options

| ID | Spark Parameters | Default Value | Modified Value |
|---|---|---|---|
| P1 | Spark.executor.memory | 1024 MB (P1-A) | 1250 MB (P1-B) |
| P2 | Spark.shuffle.sort.bypassMergeThreshold | 200 (P2-A) | 400 (P2-B) |
| P3 | Spark.shuffle.compress | TRUE (P3-A) | FALSE (P3-B) |
| P4 | Spark.memory.storageFraction | 0.5 (P4-A) | 0.7 (P4-B) |
| P5 | Spark.shuffle.file.buffer | 32 KB (P5-A) | 64 KB (P5-B) |
| P6 | Spark.reducer.maxSizeInFlight | 48 m ((P6-A) | 96 m (P6-B) |
| P7 | Spark.memory.fraction | 0.6 (P7-A) | 0.8 (P7-B) |
| P8 | Spark.serializer.objectStreamReset | 100 (P8-A) | -1 (P8-B) |
| P9 | Spark.rdd.compress | FALSE (P9-A) | TRUE (P9-B) |

Table 6: Combination of Parameter Values (CPV) executed at runtime for identifying CPV with scalability score above the threshold. 'A' and 'B' specify the default and modified value respectively

| CPV ID | Combination of Parameter Values (CPV) | P1 | P2 | P3 | P4 | P5 | P6 | P7 | P8 | P9 |
|---|---|---|---|---|---|---|---|---|---|---|
| 1 | {A, A, A, A, A, A, A, A, A} | A | A | A | A | A | A | A | A | A |
| 2 | {B, A, A, A, A, A, A, A, A} | B | A | A | A | A | A | A | A | A |
| 3 | {A, B, A, A, A, A, A, A, A} | A | B | A | A | A | A | A | A | A |
| . | . | . | . | . | . | . | . | . | . | . |
| . | . | . | . | . | . | . | . | . | . | . |
| . | . | . | . | . | . | . | . | . | . | . |
| 512 | {B, B, B, B, B, B, B, B, B} | B | B | B | B | B | B | B | B | B |

Since our approach considers a total of nine parameters each with two possible values, there are a total of 512 ($2^9$) CPVs. The execution of these many CPVs to find the CPV with a scalability score that is above the threshold is a computationally expensive task. The time required to execute and search through the large search space of 512 potential CPVs will outweigh the gain expected through adaptation. It is also worth noting that the state-of-the-art tuning approaches (e.g., [49, 50]) do follow the strategy of searching through the entire search space. However, it is computationally feasible for these approaches, which aim to tune for optimizing response time. Calculating response time requires a system to be executed only once but for calculating scalability score as required for our approach, a system needs to be executed at least five times with a different number of computing nodes. We, therefore, employed the following optimization techniques to reduce the computational time with minimal impact on the accuracy of our approach.

*Eliminating CPVs with negative Trend:* Before calculating the scalability score of a CPV for an entire curve obtained through executing the system with 1, 2, 4, 6, and 8 worker nodes, we calculate the scalability trend (Eq. 4) from six to eight nodes. If the trend is negative, then the response time increases as the cluster size changes from six to eight nodes. This implies that the CPV is not a candidate for the potential CPV with a scalability score above the threshold. The reason we include merely the transition from 6 to 8 nodes in Eq. 4 is the time overhead. To illustrate the impact of including more transitions in Eq. 4 on the time to calculate Trend, we take two sample cases - (a) using the only transition from 6 to 8 nodes and (b) including two transitions, i.e., from 4 to 6 nodes and from 6 to 8 nodes in Eq. 4. We then apply the two sample cases to the hypothetical scenarios presented in Figure 2. On average, the time required to calculate Trend in case (a) and case (b) is 9.24 sec and 14.9 sec respectively. Hence, the time required to calculate the trend in case (a) is 38.12% less than the time required to calculate the trend in case (b). This difference in time required to calculate trend increases with an increase in including more transitions (e.g., from 1 to 2 and 2 to 4 nodes) in Eq. 4. Furthermore, the results presented in Section 4.2 show that the region from 6 to 8 nodes is the most accurate region to determine if the scalability curve is showing any unexpected variation (see Section 4.2 for details). Consequently, other transitions (e.g., from 4 to 6 nodes) in Eq. 4 will have minimal impact on accuracy but a far significant impact on the time required to calculate Trend. Hence, Eq. 4 only considers the transition from 6 to 8 nodes for calculating Trend. *Keeping Change of Parameter Value with a Positive Impact:* If changing value of a parameter improves the scalability score, the changed value is kept the same for the next CPV. For example, CPV 2 achieves a better scalability score than CPV 1 by changing the value of *spark.executor.memory* from 1024 MB to 1250 MB. However, since the scalability score of CPV 2 is not above the threshold value, our algorithm will not select CPV 2 rather it will execute CPV 3, but with the *spark.executor.memory* value of 1250 MB as it has already shown a better scalability score.

Our adaptation algorithm is presented as Algorithm 1. If the scalability score of the BDCA system with default parameter settings is below the threshold (line 5), an adaptation process is triggered. The first parameter in the default CPV is changed from its default value and then the *Trend* is calculated using Eq. 4 to investigate the trend of scalability from six nodes to eight nodes. If the *Trend* is negative (i.e. response time increases as the number of nodes increases from six to eight), the parameter value is changed to its default value (lines 17-20). On the other hand, if the *Trend* is positive (response time decreases as the number of nodes increases from six to eight), a BDCA system is executed with two and four worker nodes to get the entire scalability curve (lines 21-24). After getting the scalability curve, the scalability score is calculated for the CPV. If the scalability score is higher than the previous best scalability score, the optimal CPV is updated. Finally, the scalability score of the CPV is compared with the threshold scalability score (line 29). If the scalability score is above the threshold, the CPV is selected for the future operations of a system. The variable $R_t$ in Algorithm 1 specifies the number of times each execution is repeated. Such a repetition of execution is required to remove (any) experimental fluctuations. We set $R_t$ equal to three – indicating to repeat each execution three times. We then take the mean of the response time determined in the three executions for subsequent calculation of scalability score. It is important to note that Algorithm 1 only restricts adaptation trigger based on the predefined threshold, i.e., adaptation is triggered only if the scalability score is less than the predefined threshold. Algorithm 1 ensures that it will return a CPV with scalability score either equal or better than the previously running CPV. Algorithm 1 does not guarantee that it will always return a CPV with scalability score above the predefined threshold. However, we did not observe any such case based on the results presented in Section 4.3.

## 4. Results

In this section, we present the results from our study aimed at answering the three research questions.

*4.1. RQ1: How does a BDCA system scale with default Spark configuration settings?*

This research question investigates the very premise of the work reported in this paper, i.e., to confirm whether or not a BDCA system scales ideally with the default configuration. Ideal scalability implies that a BDCA system makes full use of the additional resource provided by scaling [51]. For instance, when the number of worker nodes is doubled, the response time of a system should reduce to half. If a BDCA system scales ideally, there would be no value added by our work.

**Classification accuracy:** Before presenting scalability findings, we first present the accuracy of our BDCA system in Table 7 for the four datasets and four ML algorithms. This is because accuracy is one of the main quality measures for a BDCA system and needs to be considered before scalability [15]. According to the accuracy presented in Table 7, our system achieves a mean accuracy of 92.7% for KDD, 73.6% for DARPA, 98.2% for CIDDS, and 92.3% for CIDIDS2017. With respect to the algorithms, our system achieves a mean accuracy of 84.75% for Naïve Bayes, 95.74% for Random Forest, 83.71% for Support Vector Machine, and 92.8% for Multilayer Perceptron (MLP). The mean accuracy of our system for the four datasets and four algorithms is 89.2%, which is a decent level of accuracy as compared to the accuracy of the state-of-the-art BDCA systems [52-57]. Table 8 shows how the accuracy of the ML/DL models varies with respect to the number of nodes in the cluster. Whilst there is no significant change in the accuracy for most of the cases, the general trend shows that the accuracy slightly decreases as the number of nodes in the cluster increases. This could be attributed to the way data is distributed among the nodes during the training and testing process. A comparatively larger number of nodes in the cluster requires the generation of larger data blocks and vice versa. Such data partitioning and distribution strategy slightly impact the accuracy as presented in Table 8.

Following the approach reported in [58], we trained and evaluated the ML algorithms in a distributed manner. In other words, the cluster consists of a total of 10 nodes in our case. Among these nodes, one acts as a master and nine act as workers. The master node distributes the process of training and testing among the nine workers, which perform the training and testing in a distributed and parallel manner. On the contrary, the same job can be performed in a centralized manner – termed centralized learning, in which the training and testing of algorithms is performed centrally on a single node instead of a cluster of nodes. In addition to distributed and centralized learning of ML algorithms, deep learning approaches have gained tremendous attention in recent times [59]. Therefore, we also incorporate a Deep Learning (DL) algorithm, Multilayer Perceptron (MLP), to assess how it performs as compared to the traditional ML algorithms. We selected MLP based on its widespread usage in the cyber security domain. The accuracy and training time of the ML and DL algorithms trained and tested in centralized and distributed manners are presented in Table 9. The mean accuracy for centralized learning, distributed learning, and deep learning is 91.2%, 83.4%, and 92.83%, respectively. The difference in accuracy is due to the way algorithms are trained based on the partitioning of the data during the training process. On the other hand, the mean training time for centralized, distributed, and deep learning is 1499 sec, 588 sec, and 2285 sec. The difference in training time is due to resource allocation. For example, centralized learning takes more time as compared to distributed learning due to the more computational capacity available in distributed learning.

Table 7: Mean accuracy achieved by our BDCA system for the four datasets and three ML algorithms

| ML Algorithm | Dataset | Precision | Recall | F-Measure | False Positive Rate | Accuracy |
|---|---|---|---|---|---|---|
| Naïve Bayes | KDD | 83.4 % | 99.2 % | 90.6 % | 6.5 % | 84.2 % |
| | DARPA | 97.4 % | 55.9 % | 71.0 % | 0.1 % | 74.1 % |
| | CIDDS | 84.6 % | 100.0 % | 91.7 % | 0.4 % | 96.6 % |
| | CICIDS2017 | 46.3 % | 27.3 % | 34.4 % | 0.5 % | 83.9 % |
| Random Forest | KDD | 99.9 % | 97.1 % | 98.5 % | 2.5 % | 97.6 % |
| | DARPA | 99.9 % | 75.1 % | 85.8 % | 0.2 % | 85.8 % |
| | CIDDS | 100.0 % | 99.5 % | 99.7 % | 0.0 % | 99.9 % |
| | CICIDS2017 | 99.6 % | 97.8 % | 98.7 % | 0.7 % | 99.6 % |
| Support Vector Machine | KDD | 97.7 % | 92.4 % | 95.4 % | 0.7 % | 92.9 % |
| | DARPA | 100.0 % | 23.5 % | 38.0 % | 7.6 % | 56.6 % |
| | CIDDS | 96.1 % | 100.0 % | 91.7 % | 0.0 % | 96.7 % |
| | CICIDS2017 | 64.5 % | 57.9 % | 61.0 % | 0.5 % | 88.5 % |
| Multilayer Perceptron | KDD | 99.5% | 95.7% | 97.6% | 1.5% | 96.3% |
| | DARPA | 97.9% | 61.4% | 75.2% | 1.5% | 78.1% |
| | CIDDS | 99.9% | 98.1% | 99.0% | 0.1% | 99.6% |
| | CICIDS2017 | 98.9% | 86.5% | 92.3% | 0.2% | 97.3% |

Table 8: Accuracy achieved by our BDCA system for the four datasets and three ML algorithms in 2, 4, 6, and 8 node cluster

| | | Number of Worker Nodes in the Cluster | | | |
|---|---|---|---|---|---|
| ML Algorithm | Dataset | 2 | 4 | 6 | 8 |
| Naïve Bayes | KDD | 89.6 % | 81.7 % | 82.5 % | 80.4 % |
| | DARPA | 78.9 % | 76.6 % | 75.4 % | 73.5 % |
| | CIDDS | 94.8 % | 98.4 % | 90.7 % | 89.7 % |
| | CICIDS2017 | 88.4 % | 84.7 % | 83.5 % | 76.8 % |
| Random Forest | KDD | 96.8 % | 96.4 % | 85.9 % | 84.7 % |
| | DARPA | 88.6 % | 89.4 % | 88.4 % | 81.7 % |
| | CIDDS | 99.9 % | 98.8 % | 98.9 % | 99.1 % |
| | CICIDS2017 | 99.9 % | 99.7 % | 99.4 % | 99.2 % |
| Support Vector Machine | KDD | 88.7 % | 87.6 % | 89.4 % | 87.4 % |
| | DARPA | 77.6 % | 68.4 % | 49.7 % | 48.8 % |
| | CIDDS | 98.1 % | 97.5 % | 96.2 % | 94.5 % |
| | CICIDS2017 | 91.4 % | 90.8 % | 90.4 % | 85.4 % |
| Multilayer Perceptron | KDD | 97.4 % | 96.5 % | 97.4 % | 94.3 % |
| | DARPA | 86.4 % | 77.8 % | 79.8 % | 77.9 % |
| | CIDDS | 99.9 % | 99.8 % | 99.9 % | 96.8 % |
| | CICIDS2017 | 98.6 % | 97.8 % | 98.7 % | 93.7 % |

Table 9: Accuracy achieved by our BDCA system with centralized learning, distributed learning, and deep learning

| | | Dataset | | | | | | | |
|---|---|---|---|---|---|---|---|---|---|
| | | KDD | | DARPA | | CIDDS | | CICIDS2017 | |
| Learning Type | Dataset | Accuracy (%) | Training Time (sec) | Accuracy (%) | Training Time (sec) | Accuracy (%) | Training Time (sec) | Accuracy (%) | Training Time (sec) |
| Centralized Learning | Naïve Bayes | 90.6 | 395 | 76.4 | 2855 | 95.7 | 2141 | 88.4 | 3377 |
| | Random Forest | 90.7 | 355 | 88.1 | 2048 | 99.6 | 2122 | 99.9 | 3741 |
| | Support Vector Machine | 88.1 | 230 | 78.9 | 104 | 99.2 | 306 | 99.0 | 314 |
| Distributed Learning | Naïve Bayes | 80.4 | 331 | 73.5 | 1853 | 89.7 | 968 | 76.8 | 1521 |
| | Random Forest | 84.7 | 245 | 81.7 | 228 | 99.1 | 265 | 99.2 | 1243 |
| | Support Vector Machine | 87.4 | 184 | 48.8 | 58 | 94.5 | 120 | 85.4 | 44 |
| Deep Learning | Multilayer Perceptron | 96.3 | 312 | 78.1 | 2978 | 99.6 | 2749 | 97.3 | 3104 |

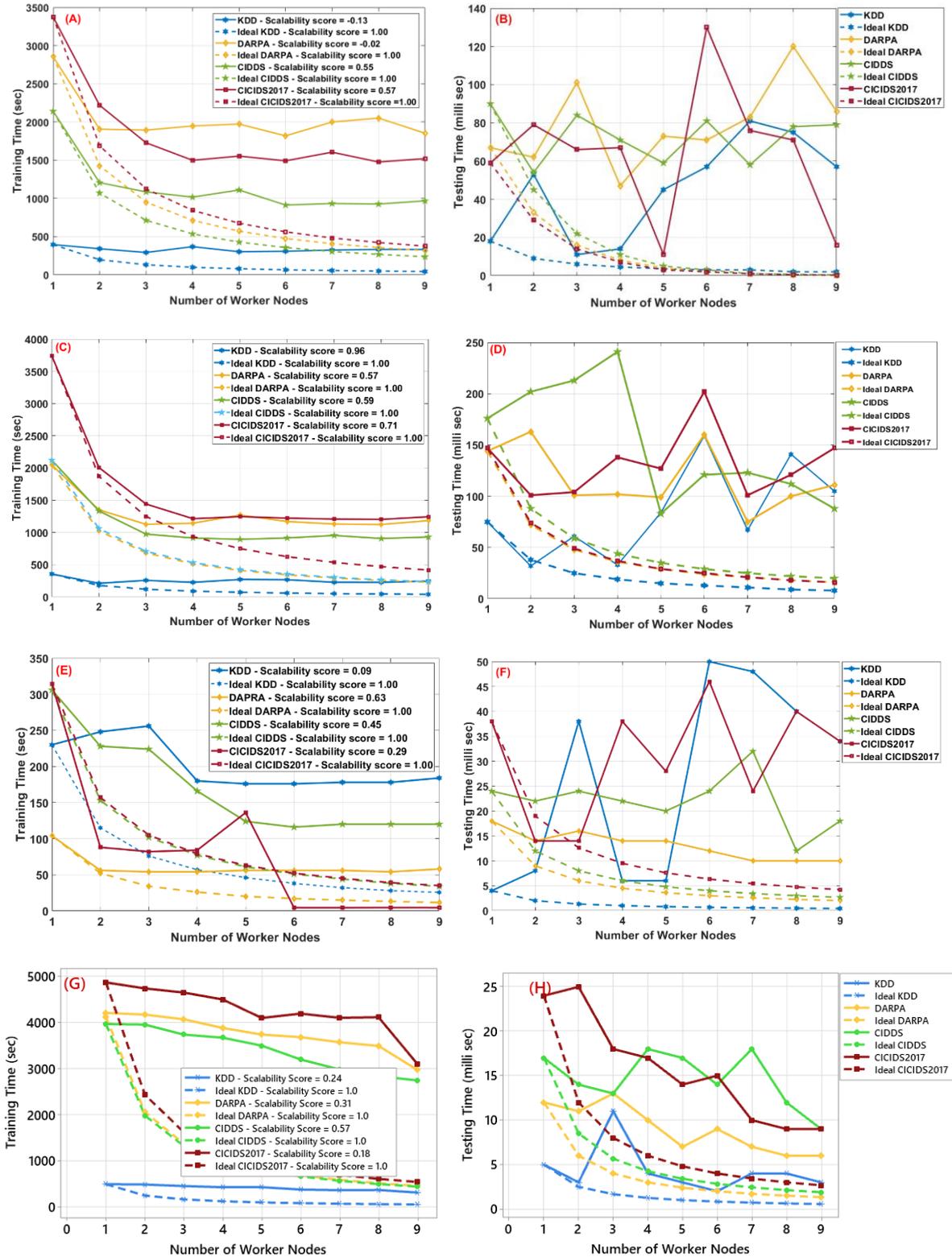

Figure 3: Ideal and achieved scalability with default Spark settings for the four datasets – KDD, DARPA, CIDDS, CICIDS2017 with (A) Naïve Bayes – training phase (B) Naïve Bayes – testing phase (C) Random Forest – training phase (D) Random Forest – testing phase (E) Support Vector Machine – training phase and (F) Support Vector Machine – testing phase. The number in the legend specifies the scalability score.

**Scalability:** Figure 3 shows the scalability curves with default Spark settings for training time and testing time of the four datasets and four algorithms. The dotted line in Figure 3 denotes ideal scalability and the solid line denotes achieved scalability. In the training phase with Naïve Bayes, Random Forest, and MLP, the system scales almost ideally as the number of worker nodes increases from one to two. The impact of adding further nodes until six nodes is negligible. After six nodes, the addition of nodes has a negative impact on scalability, i.e., the training time slightly increases. With Support Vector Machine, the trend is a bit abrupt as compared to the other three algorithms. For example, unexpected spikes can be observed at three nodes for KDD and five nodes for CICIDS2017. A potential reason for such spikes with Support Vector Machine is the short training time as compared to the training time for the other three algorithms. We used Eq. 1 to quantify the ideal and achieved scalabilities. The ideal scalability score for each dataset is **1**. The scalability scores, calculated using our scalability metric presented in Section 2.4.2, for achieved scalabilities are shown in the legend for training in Figure 3.

The mean scalability score with default Spark setting for the datasets is KDD – 0.31, DARPA – 0.39, CIDDS – 0.53, and CIDIDS2017 – 0.52. This trend is largely in line with the number of instances in each dataset. For instance, CIDDS having the largest number of instances achieves the best scalability and KDD with the smallest number of instances achieves the lowest scalability. The deviation from ideal scalability for each dataset is calculated as deviation = (1- scalability score) ×100. With regards to the algorithms, the mean scalability scores are: Naïve Bayes – 0.24, Random Forest – 0.70, Support Vector Machine – 0.36, and MLP – 0.32. The deviation from ideal scalability for each dataset is found to be; KDD – 69%, DARPA – 61%, CIDDS – 47%, and CIDIDS2017 – 48%. The deviation from ideal scalability for each algorithm is found to be; Naïve Bayes – 76%, Random Forest – 30%, Support Vector Machine – 64%, and MLP – 68%. On average, the achieved scalability of a BDCA system deviates from the ideal scalability by 59.5%. The scalability with respect to testing time is abrupt. This is because of the very quick response of the system (i.e., in milliseconds) during the testing phase. Such abrupt changes in testing time make our findings unreliable to be used for scalability analysis. Therefore, in the rest of this paper, by following the approach used in the related studies BDCA studies (e.g., [60]), we only report our findings with respect to the training time.

> ***The summary answer to RQ1:*** *A BDCA system with default Spark configuration settings does not scale ideally. The deviation from ideal scalability is around 59.5%.*

*4.2. RQ2: What is the impact of tuning Spark Configuration Parameters on the scalability of a BDCA system?*

**Impactful Spark parameters:** Table 10 shows the default values and the modified values for the 11 configuration parameters (described in Section 3) and the scalability scores achieved with the default and modified settings. Figures 4 - 7 show the scalability graphs with the default and modified values for each of the 11 parameters for the 16 use cases, i.e., 4 algorithms × 4 datasets. For instance, as shown in Table 10, changing the value of *spark.rdd.compress* from FALSE (default) to TRUE changes the scalability score of the Naïve Bayes based BDCA system from -0.14 to 0.92 for KDD, 0.53 to 0.38 for DARPA, 0.68 to 0.69 for CIDDS, and 0.61 to 0.64 for CICIDS2017. The same trend continues for the first nine parameters shown in Table 10, where modifying the value of the parameters leads to a significant change in the scalability score. The last two parameters (i.e., P10 - *spark.driver.memory* and P11 - *spark.shuffle.memoryFraction*) do not significantly impact the scalability. For example, as presented in Table 10, changing the value of *spark.shuffle.memoryFraction* from '/' (default) to 0.4 for Naïve Bayes based BDCA system brings an insignificant change in scalability score for KDD (-0.14 to -0.13), DARPA (0.53 to 0.54), CIDDS (0.68 to 0.54), and CICIDS2017 (0.61 to 0.6).

**Unexpected variations:** We also assess the regions for each of the 16 use cases (4 datasets × 4 algorithms) where unexpected variations happen. By unexpected variation, we mean a variation (e.g., from

Table 10: Scalability score with default and modified value for the 11 studied parameters. The bold numbers indicate scalability scores lower than the default scalability score. The value of only one parameter is changed from default to modified at one time such as changing value of P1 from 1024 to 1250.

| Parameter ID | | | P1 | P2 | P3 | P4 | P5 | P6 | P7 | P8 | P9 | P10 | P11 |
|---|---|---|---|---|---|---|---|---|---|---|---|---|---|
| | | | **Modified Spark Settings** | | | | | | | | | | |
| **Spark Parameter** | | Default Spark Settings (P1 = 1024, P2 = 200, P3 = TRUE, P4 = 0.5, P5 = 32k, P6 = 48m, P7 = 0.6, P8 = 100, P9 = FALSE, P10 = \, P11 = 1024) | Spark.executor.memory | Spark.shuffle.sort.bypass MergeThreshold | Spark.shuffle.compress | Spark.memory.storageFraction | Spark.shuffle.file.buffer | Spark.reducer.maxSizeInFlight | Spark.memory.fraction | Spark.serializer.objectStreamReset | Spark.rdd.compress | Spark.shuffle.memoryFraction | Spark.driver.memory |
| Modified Value | | | 1250 | 400 | FALSE | 0.7 | 64k | 96m | 0.8 | -1 | TRUE | 0.4 | 1600 |
| Naïve Bayes | KDD | -0.14 | 0.31 | 0.01 | 0.36 | 0.50 | 0.15 | 0.68 | 0.01 | 0.71 | 0.92 | -0.13 | **-0.15** |
| | DARPA | 0.53 | 0.62 | 0.81 | 0.64 | 0.67 | **0.38** | 0.68 | **0.44** | **0.43** | **0.38** | 0.54 | **0.38** |
| | CIDDS | 0.68 | 0.72 | 0.68 | 0.82 | **0.51** | 0.68 | **0.52** | **0.59** | **0.44** | 0.69 | **0.54** | **0.53** |
| | CICIDS2017 | 0.61 | 0.61 | 0.70 | **0.49** | **0.46** | **0.46** | **0.59** | 0.61 | 0.67 | 0.64 | 0.62 | **0.60** |
| Random Forest | KDD | 0.53 | **0.51** | 0.60 | 0.55 | 0.58 | **0.41** | **0.44** | 0.59 | 0.54 | **0.29** | **0.18** | **0.48** |
| | DARPA | 0.57 | 0.59 | **0.28** | 0.75 | 0.62 | 0.57 | 0.70 | **0.32** | 0.78 | 0.66 | **0.06** | **0.52** |
| | CIDDS | 0.59 | 0.63 | 0.67 | **0.48** | **0.52** | **0.45** | 0.61 | 0.70 | **0.57** | 0.60 | **-0.07** | **0.42** |
| | CICIDS2017 | 0.71 | 0.72 | **0.27** | **0.28** | **0.48** | 0.72 | **0.58** | **0.15** | 0.70 | 0.68 | 0.58 | **0.47** |
| Support Vector Machine | KDD | 0.09 | 0.25 | 0.24 | 0.41 | 0.14 | 0.30 | 0.68 | 0.17 | 0.49 | 0.66 | **0.08** | 0.31 |
| | DARPA | 0.63 | 0.66 | **0.37** | **0.54** | 0.70 | 0.65 | **0.18** | 0.73 | 0.80 | 0.83 | **0.46** | **0.59** |
| | CIDDS | 0.45 | 0.79 | 0.58 | 0.62 | 0.63 | 0.53 | 0.49 | 0.83 | 0.76 | 0.55 | **0.30** | 0.49 |
| | CICIDS2017 | 0.60 | **0.39** | **0.54** | **0.37** | 0.64 | **0.49** | 0.69 | **0.52** | 0.65 | 0.48 | 0.44 | **0.36** |
| MLP | KDD | 0.24 | 0.31 | 0.25 | 0.55 | **0.06** | 0.42 | 0.62 | 0.31 | 0.46 | 0.49 | **0.12** | **0.09** |
| | DARPA | 0.31 | 0.40 | 0.84 | **0.15** | **0.10** | 0.51 | 0.53 | **0.26** | 0.33 | **0.21** | **0.01** | **0.05** |
| | CIDDS | 0.57 | **0.09** | **0.13** | **0.06** | **0.01** | 0.88 | **0.06** | **0.06** | **0.07** | **0.02** | **0.08** | **0.04** |
| | CICIDS2017 | 0.18 | 0.19 | 0.27 | 0.47 | **0.16** | 0.37 | **0.04** | 0.28 | **0.03** | 0.15 | 0.13 | 0.42 |

Table 11: Number of unexpected variations (i.e., where training time increases unlike expected decrease) in each of the four transitions – 1 to 2 nodes, 2 to 4 nodes, 4 to 6 nodes, and 6 to 8 nodes. The value in brackets specifies the percentage of unexpected variations calculated as the number of unexpected variations divided by the number of total variations.

| ML Algorithm | Dataset | 1 to 2 nodes | 2 to 4 nodes | 4 to 6 nodes | 6 to 8 nodes |
|---|---|---|---|---|---|
| Naïve Bayes | KDD | 0 (0%) | 9 (75.0%) | 3 (25.0%) | 5 (41.6%) |
| | DARPA | 0 (0%) | 5 (41.6%) | 6 (50.0%) | 4 (33.3%) |
| | CIDDS | 0 (0%) | 0 (0%) | 7 (58.3%) | 7 (58.3%) |
| | CICIDS2017 | 0 (0%) | 0 (0%) | 8 (66.6%) | 3 (25.0%) |
| Random Forest | KDD | 1 (8.3%) | 2 (16.6%) | 1 (8.3%) | 2 (16.6%) |
| | DARPA | 0 (0%) | 0 (0%) | 2 (16.6%) | 3 (25.0%) |
| | CIDDS | 0 (0%) | 2 (16.6%) | 1 (8.3%) | 2 (16.6%) |
| | CICIDS2017 | 0 (0%) | 1 (8.3%) | 4 (33.3%) | 3 (25.0%) |
| Support Vector Machine | KDD | 1 (8.3%) | 2 (16.6%) | 2 (16.6%) | 3 (25.0%) |
| | DARPA | 0 (0%) | 1 (8.3%) | 2 (16.6%) | 2 (16.6%) |
| | CIDDS | 0 (0%) | 1 (8.3%) | 1 (8.3%) | 3 (25.0%) |
| | CICIDS2017 | 0 (0%) | 0 (0%) | 0 (0%) | 4 (33.3%) |
| Multilayer Perceptron | KDD | 2 (16.6%) | 1 (8.3%) | 2 (16.6%) | 0 (0%) |
| | DARPA | 2 (16.6%) | 0 (0%) | 2 (16.6%) | 2 (16.6%) |
| | CIDDS | 0 (0%) | 1 (8.3%) | 2 (16.6%) | 2 (16.6%) |
| | CICIDS2017 | 2 (16.6%) | 2 (16.6%) | 1 (41.6%) | 1 (8.3%) |
| Total Number of Unexpected Variations | | 8 (5.5%) | 23 (18.7%) | 37 (30.5%) | 41 (32.0%) |

2 to 4 nodes) where the training time increases instead of decreasing. The number of unexpected variations in each of the four regions for the 16 use cases are presented in Table 11. Among the 192 scalability curves (12 parameter settings × 4 datasets × 4 algorithms), 8/192 shows unexpected variation in the region from 1 to 2 nodes, 23/192 shows unexpected variation in the region from 2 to 4 nodes, 37/192 show unexpected variation in the region from 4 to 6 nodes, and 41/192 show unexpected variation in the region from 6 to 8 nodes. This trend exhibits that the rate of unexpected variation increases with the increase in the number of nodes.

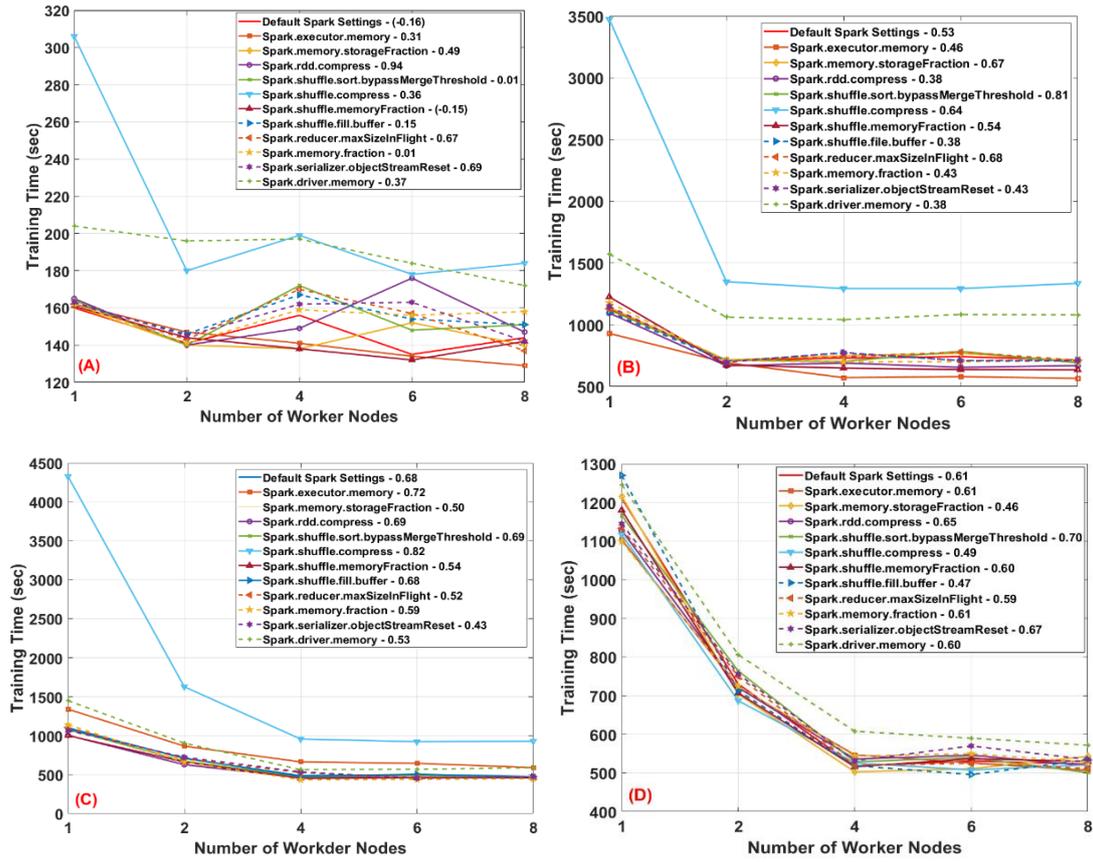

Figure 4: Impact of modifying the value of parameters on the scalability score of Naïve Bayes based BDCA system for the four datasets - (A) KDD (B) DARPA (C) CIDDS and (D) CICIDS2017. The number in the legend specifies scalability score

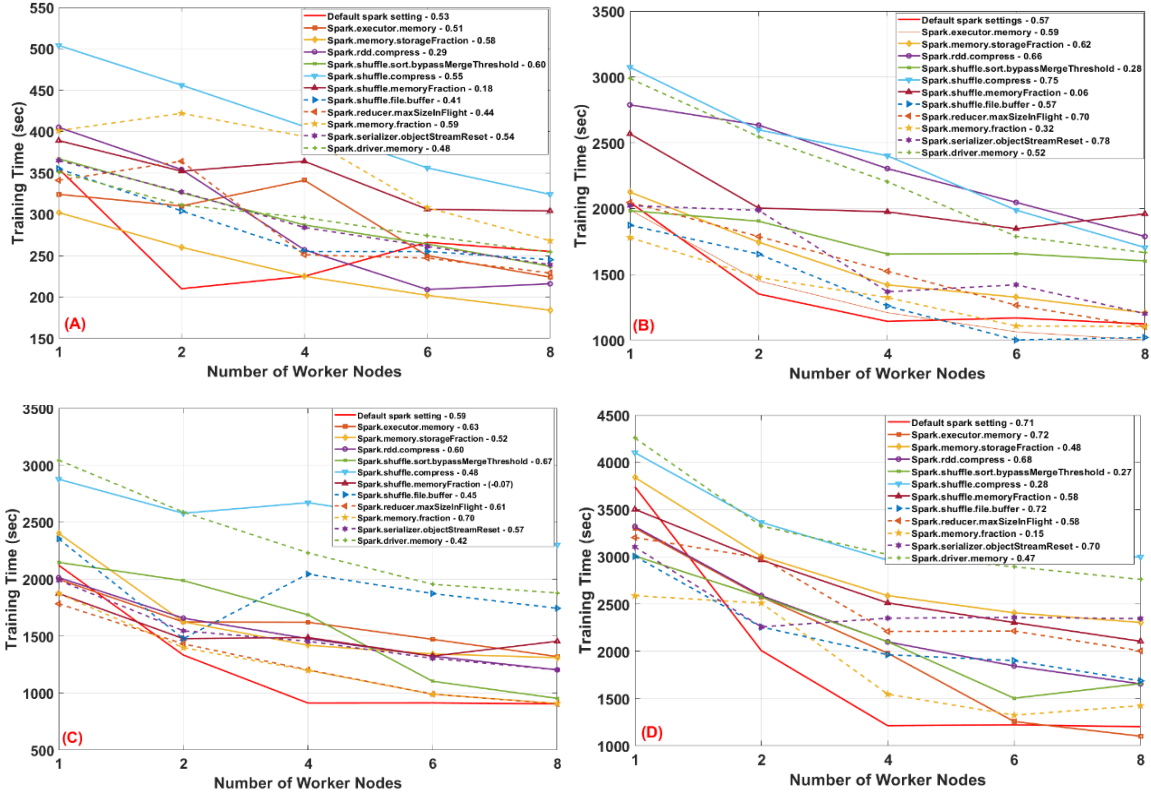

Figure 5: Impact of modifying the value of parameters on the scalability of Random Forest based BDCA system for the four datasets - (A) KDD, (B) DARPA, (C) CIDDS, and (D) CICIDS2017. The number in the legend specifies scalability score

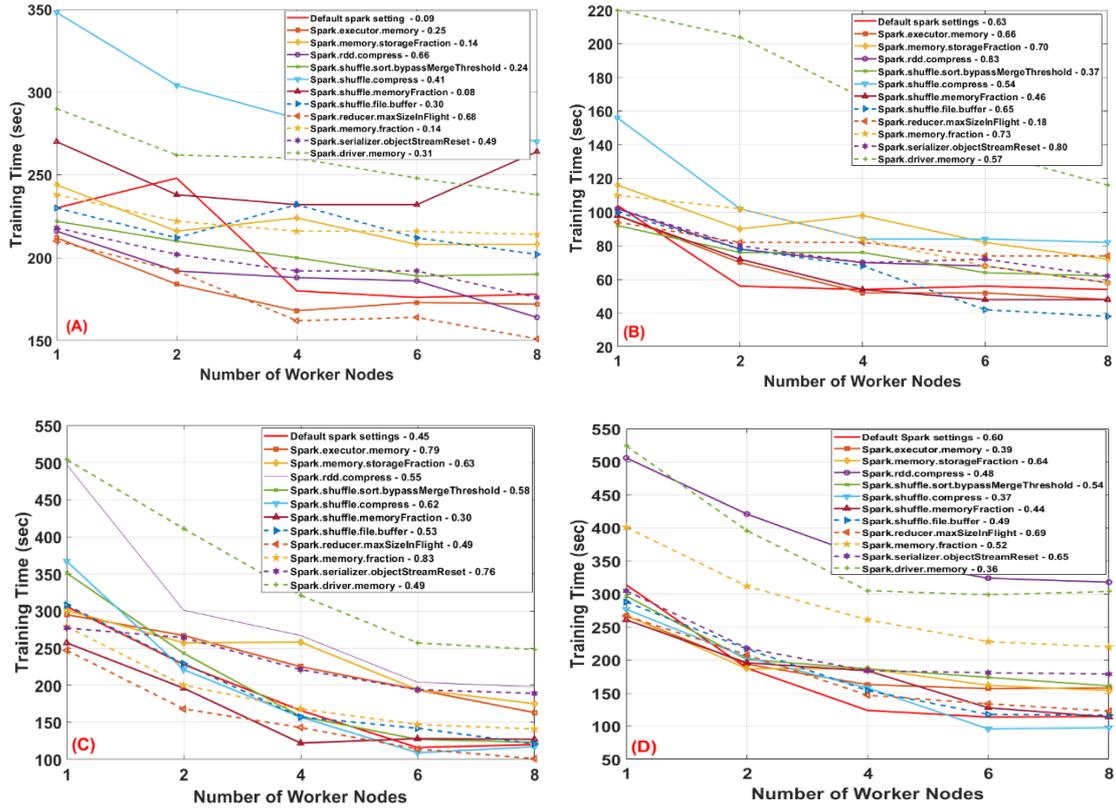

Figure 6: Impact of modifying the value of parameters on the scalability of Support Vector Machine based BDCA system for the four datasets - (A) KDD, (B) DARPA, (C) CIDDS, and (D) CICIDS2017. The number in legend specifies scalability score

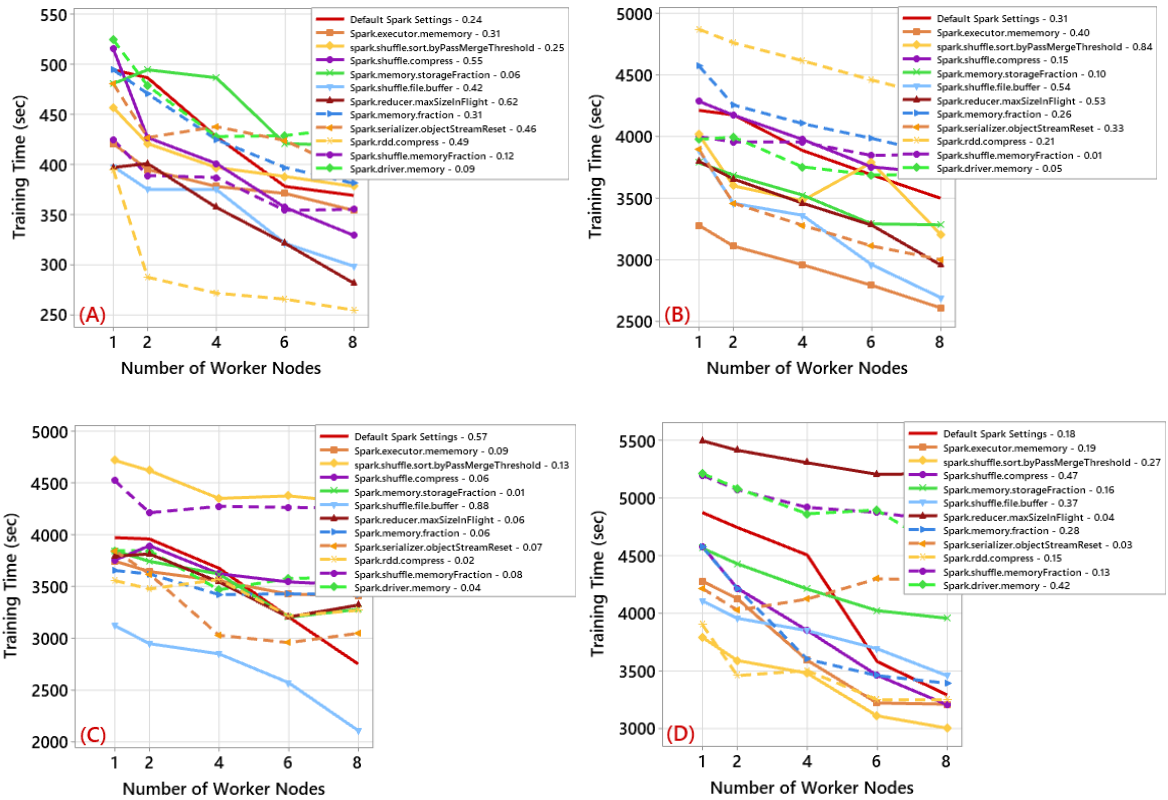

Figure 7: Impact of modifying the value of parameters on the scalability of Multilayer Perceptron based BDCA system for the four datasets - (A) KDD, (B) DARPA, (C) CIDDS, and (D) CICIDS2017. The number in legend specifies scalability

Table 12: Ranking of the studied parameters based on their impact on scalability (the number in brackets specifies the difference between the scalability score with the default settings and the modified settings)

| ID | Spark Parameters | Overall Ranking | Ranking with Respect to Datasets | | | | Ranking with Respect to ML Algorithms | | | |
|---|---|---|---|---|---|---|---|---|---|---|
| | | | KDD | DARPA | CIDDS | CICIDS2017 | Naïve Bayes | Random Forest | Support Vector Machine | Multilayer Perceptron |
| P1 | Spark.executor.memory | 9 (0.12) | 6 (0.21) | 11 (0.05) | 4 (0.14) | 9 (0.07) | 7 (0.15) | 11 (0.02) | 5 (0.19) | 11 (0.16) |
| P2 | Spark.shuffle.sort.bypassMergeThreshold | 6 (0.17) | 9 (0.12) | 1 (0.28) | 10 (0.07) | 3 (0.20) | 8 (0.13) | 3 (0.22) | 7 (0.15) | 4 (0.27) |
| P3 | Spark.shuffle.compress | 4 (0.20) | 4 (0.28) | 7 (0.13) | 6 (0.14) | 1 (0.26) | 5 (0.22) | 4 (0.19) | 4 (0.20) | 2 (0.32) |
| P4 | Spark.memory.storageFraction | 8 (0.15) | 5 (0.25) | 8 (0.09) | 5 (0.14) | 5 (0.14) | 4 (0.28) | 6 (0.10) | 11 (0.09) | 7 (0.24) |
| P5 | Spark.shuffle.file.buffer | 5 (0.16) | 7 (0.21) | 3 (0.23) | 1 (0.32) | 7 (0.09) | 6 (0.15) | 1 (0.41) | 10 (0.11) | 8 (0.23) |
| P6 | Spark.reducer.maxSizeInFlight | 1 (0.22) | 2 (0.50) | 2 (0.24) | 8 (0.07) | 8 (0.08) | 3 (0.29) | 7 (0.09) | 1 (0.29) | 1 (0.34) |
| P7 | Spark.memory.fraction | 7 (0.16) | 11 (0.09) | 6 (0.15) | 2 (0.19) | 2 (0.21) | 9 (0.08) | 2 (0.25) | 6 (0.15) | 10 (0.18) |
| P8 | Spark.serializer.objectStreamReset | 3 (0.20) | 3 (0.42) | 4 (0.16) | 3 (0.19) | 11 (0.04) | 1 (0.31) | 10 (0.06) | 3 (0.23) | 9 (0.22) |
| P9 | Spark.rdd.compress | 2 (0.21) | 1 (0.62) | 5 (0.15) | 11 (0.04) | 10 (0.06) | 2 (0.31) | 8 (0.09) | 2 (0.25) | 5 (0.24) |
| P10 | Spark.shuffle.memoryFraction | 11 (0.10) | 8 (0.12) | 10 (0.06) | 9 (0.07) | 6 (0.10) | 11 (0.04) | 9 (0.07) | 9 (0.12) | 6 (0.24) |
| P11 | Spark.driver.memory | 10 (0.10) | 10 (0.09) | 9 (0.08) | 7 (0.12) | 4 (0.16) | 10 (0.08) | 5 (0.13) | 8 (0.14) | 3 (0.30) |

**Positive/negative impact on scalability:** We assess whether modifying a parameter value has a positive or negative impact on scalability. Table 10 shows that the positive or negative impact of modifying a parameter value varies from one dataset to another as well as from one algorithm to another. The bold values in Table 10 indicate the negative impact of changing the default value for the parameter, i.e., scalability score decreases in comparison to scalability score with default settings. For example, changing the value of P3 - *spark.shuffle.compress* for Naïve Bayes based BDCA system from TRUE to FALSE has a positive impact on scalability with KDD, DARPA, and CIDDS but a negative impact on scalability with CICIDS2017. Similarly, with Random Forest based BDCA system, the default value (100) of P8 - *spark.serializer.objectStreamReset* achieves better scalability for CIDDS and CICIDS2017 while the modified value (-1) achieves better scalability for KDD and DARPA. This finding underlines a correlation between the dataset and Spark configuration parameters. We observe a similar trend for the algorithms where the optimal values of the parameters do not necessarily remain the same for different algorithms. For example, the default value of P3 - *Spark.shuffle.compress* obtains better scalability with Random Forest but the modified value achieves better scalability with Naïve Bayes and Support Vector Machine for CIDDS dataset. Hence, it can be asserted that the Spark configuration parameters need to be configured as per the type of the dataset and algorithm. In other words, this finding invalidates the reuse of a single Spark configuration setting across multiple datasets and algorithms.

**Spark parameter ranking:** Table 12 presents the ranking of the studied Spark parameters, based on their impact on scalability, with respect to the four datasets and four algorithms. The impact is calculated as the difference between the scalability score with the default Spark parameter setting and the modified one. Such a ranking is useful in prioritizing the tuning of particular parameters, i.e., parameters with a significant impact. Overall, our findings show that with respect to scalability, *Spark.reducer.maxSizeInFlight* is the most impactful and *Spark.shuffle.memoryFraction* is the least impactful Spark parameter. It is worth noting that *Spark.shuffle.compress* significantly impacts the training time as can be observed from Figure 4 – 7. However, a significant impact on training time does not necessarily mean a significant impact on scalability (Section 0). That is why it is not ranked as the most impactful with respect to scalability. Table 12 also depicts that the ranking of parameters varies with respect to datasets and algorithms. For example, *Spark.shuffle.sort.bypassMergeTreshold* is ranked as 1st and 3rd for DARPA and CICIDS2017 datasets respectively; however, this parameter is respectively ranked as 9th and 10th for KDD and CIDDS datasets. As stated earlier and illustrated by the ranking, the two parameters (i.e., *spark.driver.memory* and *spark.shuffle.memoryFraction*) are ranked at the bottom due to their insignificant or minor impact on the scalability score.

*The summary answer to RQ2:* *Modifying the default value of 9 out of 11 studied Spark parameters impacts the scalability of a BDCA system. Each security dataset and algorithm requires a separate configuration of Spark parameters for achieving optimal scalability. With respect to scalability, Spark.reducer.maxSizeInFlight is the most impactful and Spark.shuffle.memoryFraction is the least impactful Spark parameter.*

*4.3. RQ3: How to improve the scalability of a BDCA system?*

We have proposed a parameter-driven adaptation approach, *SCALER*, for improving the scalability of a BDCA system. The adaptation approach has already been described in Section 4. Here, we evaluate the effectiveness of our approach with respect to the following research questions.

*4.3.1. RQ3.1: How much scalability of a BDCA system is improved using SCALER (scalability improvement)?*

**Adaptation scenarios:** We assess the scalability improvement by comparing the scalability score achieved by our system exactly before and after adaptation. In order to realize adaptation, we experimented with two scenarios, i.e., baseline and change in input data. In the *baseline scenario*, a BDCA system is processing a particular dataset such as KDD with the optimal CPV determined for KDD based on Algorithm 1. In the *change in input data scenario*, the input to the system is changed from one dataset to another, e.g., from KDD to CIDDS. Upon the change in the dataset, *SCALER* calculates the scalability score for the new dataset (i.e., CIDDS), which is presented as the scalability score before adaptation in Table 13. If the scalability score is lower than the predefined threshold (0.58), the adaptation process is triggered. Given that we have four security datasets, a total of 12 (change in input data) use cases are possible as shown in Table 13.

**Scalability improvement:** Table 13 shows the scalability scores before and after adaptation for each of the 12 possible use cases and the mean scalability improvement for each of the four datasets. On average, *SCALER* improves scalability by 20.8%. With respect to datasets, the highest improvement is 27.83% for CIDDS followed by 25.83% for CICIDS2017, 22.71% for KDD, and 7.86% for DARPA. Since the scalability score of DARPA with Naïve Bayes is higher than the threshold score of 0.58 (Section 3.3), adaptation is not triggered for the associated three use cases. It is important to note that the scalability score after adaptation is the same for all three cases associated with each dataset. This is because *SCALER* selects a CPV for a dataset irrespective of the dataset previously being processed by the system. For example, in use cases 1 and 2, *SCALER* aims to select an optimal CPV for KDD and does not pay any attention to the previously processed datasets (i.e., DAPRA and CIDDS).

Table 13: Scalability score before and after adaptation

| Use Case ID | Use Case | Naïve Bayes | | Random Forest | | Support Vector Machine | | Multilayer Perceptron | | Mean Improvement for Dataset (%) |
|---|---|---|---|---|---|---|---|---|---|---|
| | | Scalability Score Before Adaptation | Scalability Score After Adaptation | Scalability Score Before Adaptation | Scalability Score After Adaptation | Scalability Score Before Adaptation | Scalability Score After Adaptation | Scalability Score Before Adaptation | Scalability Score After Adaptation | |
| 1 | DARPA → KDD | 0.47 | 0.63 | 0.38 | 0.61 | 0.41 | 0.59 | 0.24 | 0.61 | 26.61% |
| 2 | CIDDS → KDD | 0.52 | 0.63 | 0.49 | 0.61 | 0.47 | 0.59 | 0.38 | 0.61 | |
| 3 | CICIDS2017 → KDD | 0.33 | 0.63 | 0.54 | 0.61 | 0.52 | 0.59 | 0.51 | 0.61 | |
| 4 | KDD → DARPA | 0.70 | 0.70 | 0.50 | 0.59 | 0.60 | 0.60 | 0.59 | 0.59 | 7.61% |
| 5 | CIDDS → DARPA | 0.70 | 0.70 | 0.59 | 0.59 | 0.72 | 0.72 | 0.61 | 0.61 | |
| 6 | CICIDS2017 → DARPA | 0.70 | 0.70 | 0.41 | 0.59 | 0.54 | 0.72 | 0.54 | 0.68 | |
| 7 | KDD → CIDDS | 0.51 | 0.72 | 0.45 | 0.63 | 0.41 | 0.60 | 0.47 | 0.65 | 26.01% |
| 8 | DARPA → CIDDS | 0.54 | 0.72 | 0.63 | 0.63 | 0.53 | 0.60 | 0.55 | 0.65 | |
| 9 | CICIDS2017 → CIDDS | 0.47 | 0.72 | 0.39 | 0.63 | 0.29 | 0.60 | 0.53 | 0.65 | |
| 10 | KDD → CICIDS2017 | 0.54 | 0.60 | 0.54 | 0.71 | 0.50 | 0.64 | 0.47 | 0.64 | 22.80% |
| 11 | DARPA → CICIDS217 | 0.51 | 0.60 | 0.47 | 0.71 | 0.39 | 0.64 | 0.58 | 0.58 | |
| 12 | CIDDS → CICIDS2017 | 0.53 | 0.60 | 0.49 | 0.71 | 0.38 | 0.64 | 0.51 | 0.64 | |

Table 14: Comparison of scalability improvement achieved through *SCALER* with scalability improvement achieved with regards to the state-of-the-art approaches. The scalability improvement is calculated based on our scalability metric

| Study | Workload | Scalability Score Before Optimization | Scalability Score After Optimization | Improvement (%) | Mean Improvement (%) |
|---|---|---|---|---|---|
| Joohyun Kyong et al. [61] | Wordcount | 0.36 | 0.74 | 38.00 | 18.00% |
| | Naïve Bayes | 0.34 | 0.61 | 27.00 | |
| | Grep | 0.63 | 0.79 | 16.00 | |
| | K-means | 0.12 | 0.03 | -9.00 | |
| Hasan Jamal et al. [62] | 16KB | 0.83 | 0.86 | 3.00 | 1.75% |
| | 512KB | 0.97 | 1.0 | 3.00 | |
| | 6000KB | 0.11 | 0.13 | 2.00 | |
| | 16000KB | 0.62 | 0.61 | -1.00 | |
| Chen et al. [63] | Eclipse | 0.74 | 0.70 | -4.00 | 13.13% |
| | Hsqldb | 0.25 | 0.11 | -14.00 | |
| | Lusearch | -0.57 | 0.59 | 116 | |
| | Xalan | 0.95 | 0.79 | -16 | |
| | MolDyn | 0.75 | 0.78 | 3.00 | |
| | MonteCarlo | 0.85 | 0.80 | -5.00 | |
| | RayTracer | 0.84 | 0.85 | 1.00 | |
| | SPECjbb2005 | 0.37 | 0.61 | 24.00 | |
| *SCALER* | Naïve Bayes | 0.49 | 0.65 | 16.54 | 20.81% |
| | Random Forest | 0.44 | 0.66 | 22.50 | |
| | SVM | 0.42 | 0.63 | 23.68 | |
| | MLP | 0.49 | 0.62 | 20.34 | |

The trend in scalability improvement largely correlates with the size of each dataset – the larger the size, the larger is the improvement. A higher scalability improvement is recorded for large size datasets (i.e., CIDDS and CICIDS2017) and a lower scalability improvement is recorded for small size datasets, i.e., KDD and DARPA. With respect to algorithms, Support Vector Machine (SVM) benefits the most from *SCALER* – achieving a mean scalability improvement of 23.68%. The mean scalability improvement for Random Forest, Naïve Bayes, and MLP is 22.50%, 16.54%, and 20.3% respectively.

**Comparison with related studies:** We compare the optimization potential of *SCALER* with regards to the state-of-the-art approaches that also aim to improve the scalability of different software systems. For such a comparison, we collected the data (e.g., response time) as reported in those studies and then calculated the scalability scores, using our scalability metric (Section 2.4.2), before and after the applied optimization. The scalability score and achieved optimization in scalability for various studies are presented in Table 14. We could not make a comparison with all studies discussed in Section 6.3 due to the lack of required data in the reported studies. It is important to note the following points before we analyse the findings presented in Table 14 – (i) Since some of the studies (e.g., [61]) only report throughput, we first calculated the response time for those studies based on the reported throughput and data size (ii) some studies report findings for a cluster size greater than eight nodes. Given that our study considers a cluster size of maximum of eight nodes, we only selected (and scaled where required) the response time of up to eight nodes from those studies to make a fair comparison with those studies (iii) The studies presented in Table 14 use different datasets and different workloads. For example, Joohyun Kyong et al. [61] use BigDataBench [64] and Chen et al. [63] use DaCapo [65] in their experiments. Given that our study is focussed on security analytics, we used the datasets and algorithms used in security analytics. Therefore, owing to the usage of different datasets and algorithms in the related studies, an apple-to-apple comparison is quite challenging, and (iv) some of the studies presented in Table 14 consider various scenarios of scalability improvement. For instance, a study [61] considers two cases (e.g., fine-grained optimization and course-grained optimization). For such studies, we report the mean scalability improvement in Table 14. The improvement achieved with *SCALER* is highest among the approaches mentioned in Table 14. A potential reason for such improvement with *SCALER* is that, unlike the related studies, our approach exploits the configuration parameters of the underlying framework to improve the scalability.

> ***The summary answer to RQ3.1:*** *The proposed adaptation approach improves the scalability of the BDCA system by 20.8%. The larger the size of the dataset, the larger is the scalability improvement achieved via our proposed approach.*

*4.3.2. RQ3.2: How long does it take for SCALER to adapt a BDCA system for optimal scalability (i.e., adaptation time)?*

**Adaptation time:** The adaptation time underlines the speed with which *SCALER* adapts a system. The adaptation time is calculated as the time between the point of time adaptation is triggered to the point when the system gains a stable state, i.e., the adaptation process is terminated [48]. Figure 7 shows the adaptation time of *SCALER* for the 16 use cases, i.e., 4 datasets × 4 algorithms. On average, it takes around 170 min for *SCALER* to adapt a system, i.e., to bring a system to a state where the scalability of a system is above the predefined threshold. It is worth noting that unlike the previous studies (e.g., [66, 67]) that adapt for improving the response time, our approach takes more time due to the generation of scalability curve instead of a single point required for response time optimization. The adaptation time is mainly elapsed in executing a system with different CPVs (Table 6) to identify the CPV with which a system has a scalability score above the threshold. With respect to datasets, the mean adaptation time is longest (i.e., 374.02 min) for CICIDS2017 followed by CIDDS (133.51 min), KDD (88.03 min), and DARPA (77.75 min). This trend is largely in accordance with the size of each dataset. For example, our approach takes the longest time to adapt the BDCA system for CICIDS2017 that is the largest in size and takes the shortest time to adapt for small size datasets such as KDD and DARPA. SVM is quite fast with a mean adaptation time of 29.59 min, followed by Random Forest with a mean adaptation time of 149.15 min, Naïve Bayes with mean adaptation time of 326.25 min, and MLP with a mean adaptation time of 178 min.

**Adaptation time Vs. Training time:** The adaptation time is larger than the actual job completion time (i.e., training time). For example, the mean training time for SVM based BDCA system with the DARPA dataset is 84.33 min while the mean adaptation time for SVM based BDCA system with the DARPA dataset is 223 min. This is because in order to determine training time, a system requires to be executed only once. However, to determine the scalability score, a system needs to be executed multiple times with a different number of nodes (i.e., 1, 2, 4, 6, and 8 nodes in our case). Adaptation time is elapsed in determining scalability score for different parameter combinations; therefore, adaptation time is larger than training time. However, this factor does not invalidate the advantages of *SCALER*. Similar to most of the tuning approaches (e.g., [46, 49, 68]), the real advantage of *SCALER* is in the execution of recurring jobs (same job executed by a system multiple times over a period of time), which is a common phenomenon and equally applicable to security analytics [69, 70]. Some recent studies [69, 70] reveal that around 40% of data analytics jobs are recurrent jobs. The current job is executed for the sake of tuning; therefore, it does not benefit from tuning. However, the recurring and/or subsequent jobs benefit from the already tuned system. For example, *SCALER* improves the scalability score of a job (i.e., training SVM based BDCA system with CIDDS dataset) from 0.41 to 0.60 – an improvement of around 19%, which in turn translates into a reduction of training time from 121 min to 97.2 min with an eight nodes cluster. Now, since the system is tuned, when the system executes the same or similar job, it will take 97.2 min to complete the job instead of 121 min.

**Comparison with related studies:** In Figure 8, the number of iterations indicates the number of CPVs tried to identify the CPV with which a system has a scalability score above the threshold. On average, it takes 2.1 iterations/trails for *SCALER* to find the desired CPV from the search space. Since the related optimization approaches (e.g., [46, 49, 71, 72]) use different datasets and algorithms, we cannot make a direct comparison of the adaptation time of *SCALER* with the related approaches. However, we can make a direct comparison of the number of iterations required by each approach to converge towards a stable configuration. Such a comparison of *SCALER* with the other state-of-the-art approaches is presented in Table 15. On average, *SCALER* requires only 2.1 iterations to find the desired configuration, which is the smallest number of iterations as compared to the other state-of-the-art approaches. One of the reasons for such a small number of iterations is that instead of searching for the most optimal CPV in the search space, *SCALER* only searches for a CPV that has a scalability score above the threshold. As soon as the desired CPV is found, the search process is stopped.

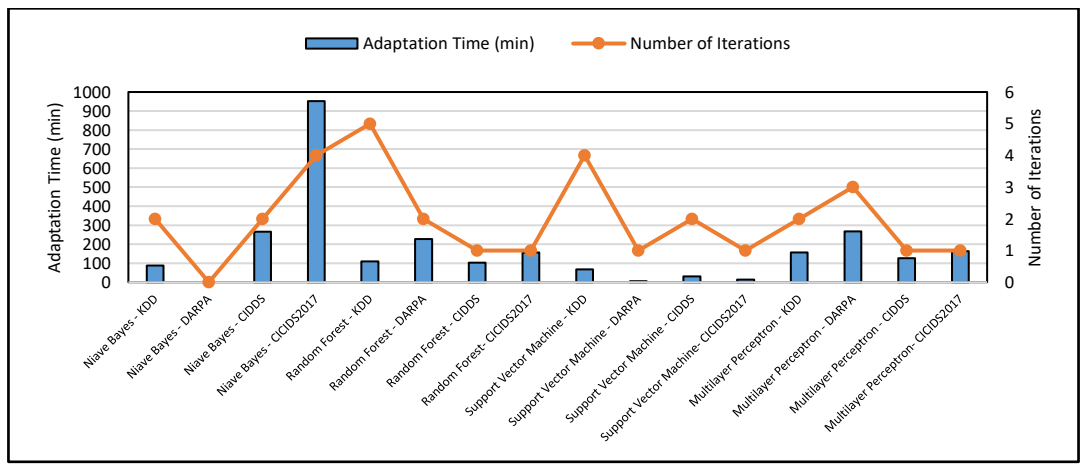

Figure 8: Adaptation time of *SCALER* for the four datasets and four ML/DL algorithms

> ***The summary answer to RQ3.2:*** *Our adaptation approach takes around 2 iterations to adapt a BDCA system to a dataset. The time taken to adapt a system is directly proportional to the size of the dataset.*

Table 15: Comparison of the number of iterations required by *SCALER* and other state-of-the-art approaches to converge towards optimal configuration

|  | Perez et al. [71] | Zhu et al. [49] | Gounaris et al. [46] | Liao et al. [72] | *SCALER* |
|---|---|---|---|---|---|
| Number of Iterations/trails to converge | 4.06 | 5 | 9 | 15.75 | 2.27 |

*4.3.3. RQ3.3: Does the number of parameters and their value options impact the optimization capability and adaptation time of SCALER?*

**Scenarios:** For this research question, we assess the impact of the number of parameters considered and their value options on the performance (i.e., scalability improvement and adaptation time) of *SCALER*. We considered four scenarios as shown in Table 16. *Scenario-1* is the default scenario, as presented in the rest of the paper, that considers nine parameters with each parameter having two potential values as shown in Table 16 and previously presented in Table 5. In *scenario-2*, we reduced the number of parameters from nine to five – considering only the most impactful parameters as determined from the average ranking presented in Table 11. *Scenario-3* considers the same nine parameters as considered in *scenario-1*, but unlike *scenario-1*, each parameter has four value options except the binary parameters such as *Spark.shuffle.compress*. The value options for the parameters are selected based on academic and industrial recommendations [32, 45]. Similarly, *scenario-4* considers the same five parameters as considered in *scenario-2*, but unlike *scenario-2*, each parameter has four value options.

**Scalability improvement:** Figure 9 (A) presents the improvement in scalability achieved by *SCALER* for each of the four studied scenarios. On average, *SCALER* improves the scalability of a BDCA system by 20.8% in *scenario-1*, 8.74% in *scenario-2*, 28.27% in *scenario-3*, and 23.62% in *scenario-4*. The improvement in scalability increases as the number of parameters and their value options increases. For example, scalability improvement is the highest (28.27%) in *scenario-3*, where a total of nine parameters are considered, each with four value options (i.e., 9 parameters – 4 value options). On the contrary, scalability improvement is the lowest (8.74%) in *scenario-2*, where *SCALER* explores combinations of only five parameters each with only two value options (i.e., 5 parameters – 2 value options). However, it is worth noting that the improvement in scalability is not directly proportional to the number of parameters and their value options considered in each scenario. For example, in *scenario-3*, *SCALER* explores almost twice the number of parameter combinations as in *scenario-1* but achieves merely 7.36% higher improvement than in *scenario-1*. A potential reason for such lack of proportionality is that Algorithm 1 does not sequentially test each parameter combination rather randomly tests parameter combinations. As a result, as soon

Table 16: Spark parameters and their value options considered in the four scenarios. The value in bold denotes the default value of the parameter

| ID | Spark Parameter | Scenario 1 | Scenario 2 | Scenario 3 | Scenario 4 |
|---|---|---|---|---|---|
| P1 | Spark.executor.memory | ✓ {**1024**, 1250} | ✗ | ✓ {**1024**, 1250, 512, 2056} | ✗ |
| P2 | Spark.shuffle.sort.bypassMergeThreshold | ✓ {**200**, 400} | ✓ {**200**, 400} | ✓ {**200**, 400, 100, 800} | ✓ {**200**, 400, 100, 800} |
| P3 | Spark.shuffle.compress | ✓ {**True**, False} | ✗ | ✓ {**True**, False} | ✗ |
| P4 | Spark.memory.storageFraction | ✓ {**0.5**, 0.7} | ✓ {**0.5**, 0.7} | ✓ {**0.5**, 0.7, 0.2, 0.9} | ✓ {**0.5**, 0.7, 0.2, 0.9} |
| P5 | Spark.shuffle.file.buffer | ✓ {**32k**, 64k} | ✓ {**32k**, 64k} | ✓ {**32k**, 64k, 16k, 128k} | ✓ {**32k**, 64k, 16k, 128k} |
| P6 | Spark.reducer.maxSizeInFlight | ✓ {**48m**, 96m} | ✓ {**48m**, 96m} | ✓ {**48m**, 96m, 24m, 192m} | ✓ {**48m**, 96m, 24m, 192m} |
| P7 | Spark.memory.fraction | ✓ {**0.6**, 0.8} | ✗ | ✓ {**0.6**, 0.8, 0.3, 1.0} | ✗ |
| P8 | Spark.serializer.objectStreamReset | ✓ {**1024**, 1250} | ✓ {**1024**, 1250} | ✓ {**1024**, 1250, 512, 2056} | ✓ {**1024**, 1250, 512, 2056} |
| P9 | Spark.rdd.compress | ✓ {**False**, True} | ✗ | ✓ {**False**, True} | ✗ |

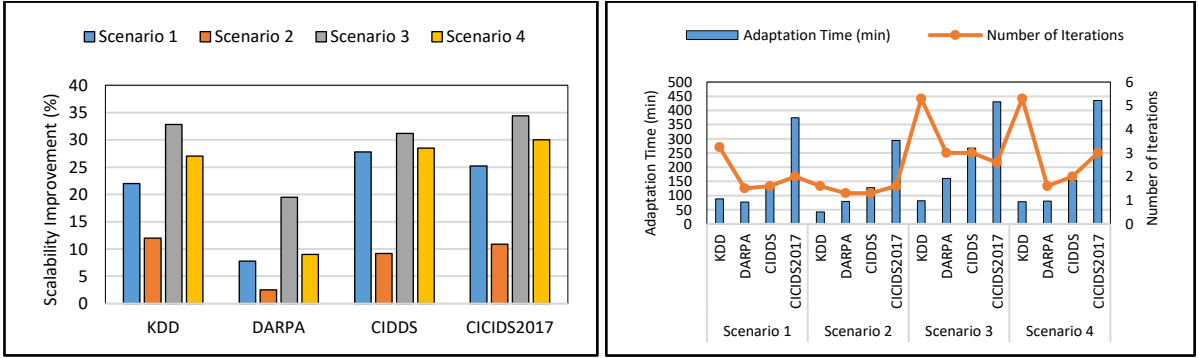

Figure 9: (A) Scalability improvement achieved by *SCALER* in each of the four scenarios presented in Table 14 and (B) Adaptation time and the number of iterations to converge towards optimal configuration

as a parameter combination with a scalability score that is above the threshold is found, the parameter combination is selected for future system operation.

**Adaptation time:** Figure 9 (B) presents the adaptation time of *SCALER* with respect to each of the four scenarios and the number of iterations to converge to a parameter combination with a scalability score above the threshold. The mean adaptation time is 168 min in *scenario-1*, 136 min in *scenario-2*, 235 min in *scenario-3*, and 187 min in *scenario-4*. As expected, the adaptation time and the number of iterations to converge towards optimal configuration increase with an increment in the number of parameters and value options. The adaptation time is the lowest (136 min) in *scenario-2,* which has the smallest number of parameter combinations (i.e., 5 parameters each with two value options). On the other hand, the adaptation time is highest (235 min) in *scenario-3,* where *SCALER* explores the highest number of parameter combinations (i.e., 9 parameters each with four value options). Although adaptation time is the lowest in *scenario-2*, *SCALER* fails to converge/find a Spark parameter setting with scalability score above the threshold in 4 out of 48 use cases (3 ML algorithms * 4 datasets * 3 changes), hence, impacting the adaptation stability [48] of *SCALER*. Such lack of convergence indicates that reducing the number of parameters in the search space may lead to a search space that does not have a parameter combination with a scalability score above the predefined threshold.

**Trade-offs:** From the above findings, we can see the case of a trade-off between optimization capability and adaptation time. Increasing the number of parameters and their value options grows the optimization capability but at the cost of increased adaptation time. For instance, in *scenario-3* that has the largest number of parameters and their value options, *SCALER* achieves the highest improvement in scalability (28.27%) but with the highest adaptation time (235 min). However, with an increment in the number of value options, the gain in optimization capability is not as significant as the loss in adaptation time. As an example, increasing the number of value options for the nine parameters from two to four (between *scenario-1* and *scenario-3*) improves the scalability by 7.36% but increases the adaptation time by 28.51%. Similarly, reducing the number of parameters from nine to five (*scenario-2* and *scenario-4*) poses a threat to the adaptation stability, i.e., *SCALER* not being able to find the desired parameter combination due to very few options. Therefore, we assert that the search space in *scenario-1* (i.e., nine parameters each with two value options) is the most suitable choice, in terms of optimization capability and adaptation time, for the applicability of *SCALER*.

> ***The summary answer to RQ3.3:*** *The higher the number of parameters and their value options in the search space, the higher is the scalability optimization with SCALER. However, a higher number of parameters and their value options in the search space also leads to higher adaptation time (trade-off between scalability improvement and adaptation time). The most suitable choice of search space is the one with nine parameters each having two value options.*

## 5. Discussion

This section discusses the broader implications of the findings.

### 5.1. Benefits to Security Operators

We now turn to the question that how the findings of this study are useful for the security operators of a BDCA system. In practice, BDCA systems are deployed and operated using default Spark settings [46]. The first takeaway from our findings is the realization that default Spark settings are not optimal from scalability perspective. Furthermore, our findings indicate that on average the deviation from ideal scalability is around 59.5%. The finding is expected to motivate a security operator to assess the scalability of a BDCA system after it is deployed. If the system scales poorly, the security operator can manually tune the parameters to improve the system's scalability. The identification and ranking of impactful parameters in Section 4.2 further facilitate a security operator to only assess the tuning of parameters that has a significant impact on a system's scalability. Our adaptation approach saves the security operator's time and effort that is to be invested in finding the right combination of parameters with which a system can scale in a better way.

### 5.2. Extending SCALER to Other Domains

Whilst the main aim of this study was to investigate and improve the scalability of a BDCA system, the techniques (e.g., *SCALER* and parameters' impact) presented in this paper can be extended and applied to other domains. One such domain is banking where big data analytics is making its mark. In banking, big data analytics is used to analyse large volumes of data to understand in (near real-time) customer behaviour, promote the right product, and increase revenue [73]. Similar to security data, the generation of banking data (e.g., transactions) increases and decreases with time [74]. For example, a heavy workload is observed during working hours and at the end of a month. Accordingly, the banking big data analytics systems need to scale as per the workload. Therefore, our adaptation approach (Section 4) can be applied to banking big data systems for automatic tuning to improve the scalability of a system. Another potential domain is healthcare analytics where big data technologies are frequently employed to deal with massive volumes of healthcare data (e.g., patient records) [75]. For healthcare big data systems, the workload fluctuates frequently as a heavy workload is to be handled in an emergency (e.g., natural disasters) [76]. Hence, we believe that a healthcare big data system can also benefit from using our adaptation approach.

### 5.3. Experimental Bottlenecks

During our experimentation, we faced several bottlenecks related to the hardware resources and Spark processing framework. We discuss those issues for the benefit of interested readers (i.e., researchers), who may come across the same issues. Our experiments produced temporary data during job execution on Spark, which consumes a lot of disk space on worker nodes. Since each worker node has a limited disk space (10 GB in our case), the temporary data can exceed the disk limit of the worker node. In such a case, the worker node becomes unhealthy and so the master node does not assign the worker node any further tasks. To deal with this issue, we used to delete the temporary data produced on worker nodes. However, we ensured that critical data such as the HDFS block files and the block metafiles are not deleted. Debugging becomes a serious concern in distributed data processing. We also initially faced the challenges of debugging failures (e.g., node failure) during our experiments. For instance, running a data processing job with eight worker nodes, where two worker nodes are already unhealthy, consumes time but produces useless results. This is because the experiment was designed for eight nodes, but two nodes were in an unhealthy state and we were not aware of it. To handle this issue, we designated a path through the variable *yarn.nodemanager.log-dirs* as the path for saving operational logs. We used to constantly check the logs to identify any issues before running the experiment.

### 5.4. Threats to Validity

In this study, we have investigated a specific BDCA system that is using particular algorithms and a big data framework (Spark). Therefore, our findings may not generalize to all kinds of BDCA systems. However, it is important to note that the aim of this study is not to show the results that generalize to all BDCA systems but to show that the parameter configuration of the underlying big data framework impacts a BDCA system's scalability. Nonetheless, future research aimed at obtaining more generalizable results will be useful. The number of value options (i.e., two and four) we investigated for the parameters limit exploration of the parameter space. Since the modification of a parameter value (e.g., from 1024 MB to 1250 MB) shows a significant impact on scalability, investigating other modifications (e.g., from 1024 MB to 2056 MB) can only strengthen our findings but cannot contradict them. Our adaptation approach takes around 170 min to select Spark configuration with a scalability score above the threshold. Although the real advantage of our approach is the reduction in the execution time of the recurring jobs, the adaptation time can be reduced in the future by (i) reducing the number of parameters considered during tuning through techniques such as Lasso linear regression [77] and (ii) similar to [68] and [78], using representative datasets of smaller size instead of using the original datasets. Our study has only investigated a limited number of parameters for their impact on scalability. Even if other Spark parameters do not impact scalability, our findings for the studied parameters still remain valid. For feature engineering, we have used *StringIndexer* (from *org.apache.spark.ml.feature*) to transform the label features (i.e., normal and attack) in the KDD dataset from string to indices as described in Section 2.2.1. However, this approach introduces order in features that do not have a natural order, which results in affecting/biasing the results of the machine learning model. Therefore, using the ordinal encoding of string features limits our approach and consequently poses a threat to the validity of our findings related to the KDD dataset. In the future, it will be interesting to incorporate and assess other encoding techniques, such as one-hot encoding, to discard such bias and make a comparison with the existing results.

# 6. Related Work

In this section, we compare our study with the existing studies on BDCA systems, scalability investigation, scalability optimization, and adaptation approaches.

6.1. BDCA Systems

Given the exponentially growing number of cyber-attacks and increasing emphasis on real (or near)-time cybersecurity data analytics, there is strong interest in the strategies and tools of engineering and operating optimal BDCA systems. However, there is relatively less literature on this topic [15]. Spark-based BDCA systems are rapidly surpassing Hadoop-based BDCA systems in popularity and adoption as 70% of the BDCA systems in 2014 were Hadoop-based and only 30% were Spark-based; it changed to 50% Hadoop-based and 50% Spark-based in 2017 [15]. Recently several studies (e.g., [52-57]) have proposed Spark-based BDCA systems. Gupta et al. [52] present a Spark-based BDCA system that leverages two feature selection algorithms (i.e., correlation-based feature selection and Chi-squared feature selection) and several ML algorithms for detecting cyber intrusions. The system was evaluated with KDD dataset. The Spark-based BDCA system presented in [53] used K-means clustering for intrusion detection. Marchal et al. [54] propose a Spark-based BDCA system for collecting different types of security data (e.g., HTTP, DNS, IP Flow, etc.) and correlating the data to detect cyber-attacks. Las-Kasas et al. [55] present a Spark-based BDCA system that leverages Apache Pig, Apache Hive, and SparkSQL to collect emails from honeypots installed in different countries and analyse the emails to detect phishing attacks. Another Spark-based BDCA system presented in [56] analyses abnormal network packets for unveiling DoS attacks. RADISH [57] is another Spark-based BDCA system that aims to detect abnormal user and resource behaviour in an enterprise to detect insider threats. Similarly, Wang et al. [79] focussed on the 3 Vs (volume, variety, and veracity) of cyber security big data to explore the impact of missing values, duplicates, variable correlation, and general data quality on the detection of cyber-attacks. The authors used R language and several datasets such as KDD-Cup and MAWILab in their study. Like the previous studies, our study also uses Spark and ML algorithms for detecting cyber intrusions. However, unlike the previous studies, our study has been evaluated with four ML/DL algorithms and four different security datasets in a fully distributed mode, which enables us to assert that our findings are based on a more rigorous study and are more generalizable. Since the previous studies use different ML algorithms and security datasets for evaluation, hence, an apple-to-apple comparison of our findings with the findings from the previous studies is not possible.

6.2. Scalability of BDCA Systems

Despite the increasing importance of the scalability of BDCA systems as reported in several studies [15], there have been only a few efforts (e.g., [18, 55, 60, 80, 81]) aimed at investigating the scalability of BDCA systems. Lee et al. [18] investigated the scalability of a Hadoop-based BDCA system on a 30-node cluster and observed that the execution time improves in proportion to the hardware resource from 5 to 30 nodes. Du et al. [80] studied the scalability of a Storm-based BDCA system on a five-node cluster and observed that the system failed to achieve an ideal level of scalability due to extra task scheduling and communications overhead between *spout* and *bold* phases of the Storm execution environment. Aljarah et al. [60] also studied the scalability of a Hadoop-based BDCA system on an 18 node cluster and found that the system scaled abruptly as the number of nodes in the cluster was increased. For example, an ideal speedup is observed from two to four nodes and 14 to 16 nodes while the non-ideal speedup is observed for the rest of the scalability curve. The non-ideal speedup is attributed to the start-up of MapReduce jobs and storing intermediate results in HDFS. Las-Casas et al., [55] compared the scalability of two BDCA systems – one Hadoop-based and another Spark-based. They found that Spark scales better than Hadoop due to the efficient use of caching in Spark. Xiang et al. [81] also explored the scalability of a BDCA system on a 30-node cluster and found that the execution time decreases, although not ideally, with an increase in the number of nodes up to 25 nodes. After 25 nodes, the execution time is increased, which the authors attribute to the excessive communication among nodes and disk read/write operation during MapReduce tasks. Whilst the previous studies have investigated the scalability of a BDCA system, none of the studies have quantified the scalability; nor have they calculated the deviation from the ideal scalability. Furthermore, the previous studies have only investigated the scalability with default settings. Our study is the first study that has (i) quantified the scalability with respect to four datasets (ii) assessed the deviation from the ideal scalability and most importantly (iii) investigated the impact of Spark parameters on the scalability of a BDCA system.

6.3. Scalability Improvement

Several studies (e.g., [61-63, 82-84]) have proposed methods for improving the scalability of software systems in different domains. Kyong et al. [61] proposed a docker container-based architecture for Spark-based scale-up server, where the original scale-up server is partitioned into several small servers to reduce memory access overheads. Wu et al. [82] propose a scalability improvement technique that learns from the interaction patterns among the services of a service-based application and accordingly adopts an optimized task assignment strategy to reduce the communication bandwidth and improve the scalability. Senger et al. [83] defined a scalability measure called input file affinity that quantifies the level of file sharing among tasks belonging to a bag-of-tasks application (e.g., data mining algorithm). In a study [83], the authors proposed a scalability improvement method that leverages the input file affinity measure to increase the degree of file sharing among tasks. Chen et al. [63] first studied the scalability of Java applications with default JVM settings and then proposed a tuning approach that alleviates JVM bottlenecks to improve the scalability of Java applications. Hasan et al. [62] studied the scalability of virtual machine-based systems on a multicore processor setup. This study revealed that excessive communication among virtual machines impacts the scalability of multicore systems. Canali et al., [84] present a scalability improvement approach for cloud-based systems, which leverages the resource usage patterns (e.g., CPU, storage, and network) among virtual machines and accordingly group the virtual machines in a cloud-based infrastructure. It is important to note that some studies [46, 47, 71, 85] proposed tuning techniques for Spark-based systems with the objective to reduce execution time. Such studies are largely impertinent to ours as they are focussed on execution time (response time). Our study focusses on scalability, which are two very different quality attributes (i.e., response time and scalability) of a software system and are treated differently in the state-of-the-art [86]. Therefore, the approaches presented in [46, 47, 71, 85] are not aimed at improving scalability. In general, the previous studies are largely orthogonal to our study. This is because (i) our study is the first of its kind that aims to improve the scalability of Spark-based BDCA systems and (ii) employ a parameter-driven adaptation approach for improving the scalability of a BDCA system that, unlike the previous studies, automatically improves scalability at runtime.

6.4. Parameter-driven Adaptation

Parameter-driven adaptation is one of the commonly used adaptation approaches. Several studies (e.g., [87, 88] [89, 90]) attempted to modify the values of a certain system's parameters to achieve various objectives such as high accuracy and improved security. Calinescu et al. [87] used the KAMI model based on a Bayesian estimator to modify model parameters at runtime for achieving reliability and quick response in a service-based medical assistance system. Another study [88] argued that the software abstraction models' parameters such as Discrete-Time Markov Chains (DTMC) should be constantly updated to achieve better accuracy. The authors of the study [88] proposed an adaptation method that leverages real-time operational data of a system to keep the parameters up to date. Similarly, parameter-based adaptation is quite common in the ML domain for adjusting a model's parameters. For example, Tongchim et al. [89] proposed a parameter-driven adaptation approach for adjusting the control parameters of genetic algorithms to achieve optimal accuracy. The approach reported in [89] divides the parameter space into sub-spaces and each sub-space evolves on separate computing nodes in parallel. Jiang et al. [90] proposed a parameter-driven adaptation approach that uses the temporal and spatial correlations among characteristics (such as size and velocity of objects) for finding the best set of configuration parameters for a convolutional neural network employed in a video analytics system. The adaptation approach proposed in [90] aims to balance resource consumption and a system's accuracy. From the adaptation point of view, our study differs from the previous studies in two ways. First, our study is the first of its kind that applies a parameter-driven adaptation approach in the domain of Spark-based systems. Second, unlike the previous studies that aim to achieve accuracy or quick response time, our adaptation approach aims to achieve improved scalability.

## 7. Conclusion

Big Data Cyber Security Analytics (BDCA) systems use big data technologies (such as Apache Spark) to collect and analyse security event data (e.g., NetFlow) for detecting cyber-attacks such as SQL injection and brute force. The exponential growth in the volume and the unpredictable velocity of security event data require BDCA systems to be highly scalable. Therefore, in this paper, we have studied (i) how a Spark-based BDCA system scales with default Spark settings (ii) how tuning configuration parameters (e.g., execution memory) of Spark impacts the scalability of a BDCA system, and (iii) proposed *SCALER* - a parameter-driven adaptation approach to improve a BDCA system's scalability. For this study, we have developed an experimental infrastructure using a large-scale OpenStack cloud. We have implemented a Spark-based BDCA system and have used four security datasets to find out how a BDCA system scales, how Spark parameters impact scalability, and to evaluate our adaptation approach aimed at improving scalability. Based on our detailed experiments, we have found that:

- With default Spark settings, a BDCA system does not scale ideally. The deviation from ideal scalability is around 59.5%. The system scales better with large size datasets (e.g., CICIDS2017) as compared to small size datasets (e.g., KDD)
- 9 out of 11 studied Spark parameters impact the scalability of a BDCA system. The impact of configuration parameters on scalability varies from one security dataset to another
- Our parameter-driven adaptation approach improves the mean scalability of a BDCA system by 20.8%.

From our findings, we conclude that practitioners should first tune the parameters of Spark before putting a Spark-based BDCA system into operation. Such parameter tunning can improve the scalability of the system. We also recommend that practitioners should not use someone else's pre-tuned parameter settings. The reason for this is that the best combination of Spark parameters varies from dataset to dataset. Our proposed adaptation approach is the first step towards facilitating practitioners to automatically tune Spark parameters for achieving optimal scalability. More generally, we assert that the field of big data analytics should pay attention to the impact of the configuration parameters of the big data frameworks on the various system qualities such as reliability, response time, and scalability. Federated machine learning has recently gained tremendous attention in various domains due to its ability to perform on-device collaborative training in a privacy-preserved manner [91]. It'd be worth exploring how our proposed approach performs with respect to federated machine learning.

Based on our study, we highlight the following areas for future research. *Investigating the parameters' impact of other big data frameworks*. Although currently, Spark is the most popular big data framework, there exist several other big data frameworks (such as Hadoop [28], Storm [29], Samza [30], and Flink [92]) with a different set of configuration parameters. Therefore, future research should investigate how configuration parameters of these frameworks impact the scalability of a BDCA system. *Approximate analytics for tuning big data frameworks:* Approximate analytics is an emerging concept that encourages computing over a representative sample instead of computing over the entire dataset [93]. The rationale behind approximate analytics is to make a trade-off between accuracy and computational time. In our study, we used the entire security datasets for system execution and subsequent tuning. Therefore, an interesting avenue for future research is to explore the applicability of approximate analytics for tuning big data frameworks. *Investigating the parameters' impact on other system's qualities*. The focus of our study was only on scalability, however, there exist several other quality attributes (e.g., reliability, security, and interoperability) that are also important for a BDCA system. Hence, it is worth investigating that how the configuration parameters of Spark impact other quality attributes of a BDCA system.

**Acknowledgement**

The authors would like to thank Anying Xiang for her help in conducting the experiments.